\documentclass[11pt,a4paper]{article}

\usepackage{amssymb}
\usepackage[dvips]{graphicx}
\usepackage{bm}

\begin{document}

\title{\bf Exact renormalization of the photino mass in softly broken ${\cal N}=1$ SQED with $N_f$ flavors regularized by higher derivatives}

\author{
I.V.Nartsev, K.V.Stepanyantz\\
{\small{\em Moscow State University}}, {\small{\em Faculty of Physics, Department  of Theoretical Physics}}\\
{\small{\em 119991, Moscow, Russia}}}

\maketitle

\begin{abstract}
We consider the softly broken ${\cal N}=1$ supersymmetric electrodynamics, regularized by higher derivatives. For this theory we demonstrate that the renormalization of the photino mass is determined by integrals of double total derivatives in the momentum space in all orders. Consequently, it is possible to derive the NSVZ-like exact relation between the photino mass anomalous dimension and the anomalous dimension of the matter superfields in the rigid theory by direct summation of supergraphs. It is important that both these renormalization group functions are defined in terms of the bare coupling constant, so that the considered NSVZ-like relation is valid independently of the subtraction scheme in the case of using the higher derivative regularization. The factorization of integrals defining the photino mass renormalization into integrals of double total derivatives is verified by an explicit two-loop calculation.
\end{abstract}

\unitlength=1cm

\section{Introduction}
\hspace*{\parindent}

An interesting feature of ${\cal N}=1$ supersymmetric gauge theories is the existence of the relation between the $\beta$-function and the anomalous dimension of the matter superfields, which is called "the exact NSVZ $\beta$-function" \cite{Novikov:1983uc,Jones:1983ip,Novikov:1985rd,Shifman:1986zi}. (For the pure Yang--Mills theory it gives the exact $\beta$-function in the form of a geometric series.) A similar relation also exists in softly broken ${\cal N}=1$ supersymmetric theories for the anomalous dimension of the gaugino mass \cite{Hisano:1997ua,Jack:1997pa,Avdeev:1997vx}. This relation can be presented in the form \cite{Hisano:1997ua}

\begin{equation}\label{Hisano_Shifman_Relation}
\frac{\alpha m}{\beta(\alpha)} = \mbox{RGI},
\end{equation}

\noindent
where $\beta(\alpha)$ in Eq. (\ref{Hisano_Shifman_Relation}) is the exact NSVZ $\beta$-function for the considered theory, $m$ is the gaugino mass, and $\mbox{RGI}$ denotes that this expression is the renormalization group (RG) invariant.

In this paper we will discuss the softly broken ${\cal N}=1$ supersymmetric electrodynamics (SQED) with $N_f$ flavors, for which the exact NSVZ $\beta$-function is written as \cite{Vainshtein:1986ja,Shifman:1985fi}

\begin{equation}\label{NSVZ_Beta_Renormalized}
\beta(\alpha) = \frac{\alpha^2 N_f}{\pi}\Big(1-\gamma(\alpha)\Big)
\end{equation}

\noindent
and relates the $\beta$-function to the anomalous dimension of the matter superfields $\gamma(\alpha)$. Differentiating Eq. (\ref{Hisano_Shifman_Relation}) with respect to $\ln \mu$, where $\mu$ is the renormalization point, after some transformations one obtains

\begin{equation}\label{M_RG_Equation}
\frac{d}{d\ln\mu}\Big(\frac{m}{\alpha}\Big) = - \frac{2m\beta(\alpha)}{\alpha^2} + \frac{m}{\alpha} \frac{d\beta(\alpha)}{d\alpha}
= m\alpha \frac{d}{d\alpha}\Big(\frac{\beta(\alpha)}{\alpha^2}\Big).
\end{equation}

\noindent
This all-order equation (in which $\beta(\alpha)$ is not necessarily coincides with the NSVZ expression) and its non-Abelian generalization can be obtained by using analytic continuation of the coupling constant into superspace \cite{ArkaniHamed:1998kj}. Namely, if we are interested only in the renormalization of the photino mass, it is possible to write the renormalization group equation in terms of the coupling constant superfield \cite{Hisano:1997ua}

\begin{equation}\label{Coupling_Superfield}
A \equiv \alpha(1+m\theta^2)
\end{equation}

\noindent
in the form

\begin{equation}
\frac{dA}{d\ln\mu} = \beta(A).
\end{equation}

\noindent
Then the terms without $\theta^2$ give the usual renormalization group (RG) equation for the coupling constant, while the terms proportional to $\theta^2$ automatically give Eq. (\ref{M_RG_Equation}).\footnote{Note that one can define the real coupling constant superfield depending on both $\theta^2$ and $\bar\theta^2$ \cite{ArkaniHamed:1998kj,Terao:2001jw,Kazakov:2002bt}. However, for investigating renormalization of the photino mass the terms containing $\bar\theta^2$ are not essential. The coupling constant (\ref{Coupling_Superfield}) is obtained from the real coupling constant superfield by the formal substitutions $\bar\theta^2\to 0$.}
Therefore, Eq. (\ref{M_RG_Equation}) is obtained for arbitrary subtractions respecting the superfield structure (\ref{Coupling_Superfield}).

Substituting the exact NSVZ $\beta$-function (\ref{NSVZ_Beta_Renormalized}) into the equation (\ref{M_RG_Equation}) we relate the expression in the left hand side to the anomalous dimension of the matter superfields:

\begin{equation}\label{Soft_Mass_Renormalized}
\frac{d}{d\ln\mu}\Big(\frac{m}{\alpha}\Big) = -\frac{m\alpha N_f}{\pi}\cdot \frac{d\gamma(\alpha)}{d\alpha}.
\end{equation}

\noindent
Note that unlike Eqs. (\ref{Hisano_Shifman_Relation}) and (\ref{M_RG_Equation}) this equation is not automatically obtained if the coupling constant is treated as the superfield. The matter is that deriving Eq. (\ref{Soft_Mass_Renormalized}) it is necessary to involve the NSVZ relation which is valid only in a certain class of subtraction schemes. In terms of the gauge coupling superfield (\ref{Coupling_Superfield}) Eq. (\ref{Soft_Mass_Renormalized}) can be considered as the $\theta^2$-component of the superfield NSVZ relation

\begin{equation}\label{Superfield_NSVZ}
\beta(A) = \frac{A^2 N_f}{\pi}\Big(1-\gamma(A)\Big).
\end{equation}

\noindent
However, the NSVZ relation was derived from general arguments (see, e.g., \cite{Shifman:1999mv,ArkaniHamed:1997mj,Kraus:2002nu}) and, up to now, it is not completely clear in which subtraction scheme it is obtained. Note that according to the explicit three-loop calculations the NSVZ scheme does not coincide with the MOM scheme \cite{Kataev:2013csa,Kataev:2014gxa} and with the $\overline{\mbox{DR}}$ scheme \cite{Jack:1996vg,Jack:1996cn,Jack:1998uj}. Nevertheless, it can be related to each of these schemes by a finite renormalization. The subtraction scheme in which Eq. (\ref{Superfield_NSVZ}) is valid has not also been found. (Below in this paper we demonstrate that the superfield equation (\ref{Superfield_NSVZ}) depends on the subtraction scheme similarly to the ordinary NSVZ relation.) However, the NSVZ scheme has been
constructed for (rigid) ${\cal N}=1$ SQED with $N_f$ flavors \cite{Kataev:2013csa,Kataev:2014gxa,Kataev:2013eta}. In order to specify this subtraction scheme one should
regularize the theory by higher derivatives \cite{Slavnov:1971aw,Slavnov:1972sq}.\footnote{By construction, this regularization includes inserting the Pauli--Villars determinants which cancel the one-loop divergences \cite{Slavnov:1977zf}.} (Unlike the dimensional reduction \cite{Siegel:1979wq,Siegel:1980qs}, this method is
mathematically consistent and can be formulated in a manifestly ${\cal N}=1$ supersymmetric way \cite{Krivoshchekov:1978xg,West:1985jx}. ${\cal N}=2$ generalizations are also possible \cite{Krivoshchekov:1985pq,Buchbinder:2014wra,Buchbinder:2015eva}. For investigating the softly broken ${\cal N}=1$ supersymmetric theories this method was applied in \cite{ArkaniHamed:1998kj}.) The main observation which allows constructing the NSVZ scheme is that, in the case of using the
higher derivative regularization, the RG functions of ${\cal N}=1$ SQED with $N_f$ flavors defined in terms of the bare coupling constant (see Eq. (\ref{Bare_Definitions})
below) satisfy the NSVZ relation

\begin{equation}\label{NSVZ_Beta_Bare}
\beta(\alpha_0) = \frac{\alpha_0^2 N_f}{\pi}\Big(1-\gamma(\alpha_0)\Big)
\end{equation}

\noindent
in all orders independently of the subtraction scheme \cite{Stepanyantz:2011jy,Stepanyantz:2014ima}. The scheme independence follows from the fact \cite{Kataev:2013eta} that these RG functions (defined in terms of the bare coupling constant) depend on a regularization, but do not depend on a subtraction scheme if a regularization is fixed.
Eq. (\ref{NSVZ_Beta_Bare}) follows from the factorization of integrals defining $\beta(\alpha_0)$ into integrals of (double) total derivatives \cite{Soloshenko:2003nc,Smilga:2004zr}. This feature of quantum corrections has been rigorously proved in all orders in \cite{Stepanyantz:2011jy,Stepanyantz:2014ima} and was confirmed by explicit three-loop calculations \cite{Kazantsev:2014yna}. Factorization into double total derivatives was also demonstrated for various non-Abelian supersymmetric theories and for various versions of the higher derivative regularization \cite{Pimenov:2009hv,Stepanyantz_MIAN,Stepanyantz:2011bz,Stepanyantz:2012zz,Stepanyantz:2012us,Aleshin:2016yvj,Buchbinder:2014wra,Buchbinder:2015eva} in the lowest orders of the perturbation theory.

The scheme-dependence becomes essential for the RG functions defined (standardly) in terms of the renormalized coupling constant \cite{Bogolyubov:1980nc}. Starting from Eq. (\ref{NSVZ_Beta_Bare}), it is possible to construct a simple prescription giving the subtraction scheme in which the RG functions defined in terms of the renormalized coupling constant satisfy the NSVZ relation in all orders \cite{Kataev:2013csa,Kataev:2014gxa,Kataev:2013eta}, if the (Abelian) supersymmetric theory is regularized by higher derivatives.\footnote{A possible form of a similar prescription defining the NSVZ scheme for the non-Abelian supersymmetric theories regularized by higher covariant derivatives was discussed in \cite{Stepanyantz:2016gtk}.} Up to now, no analogs of this result have been found in the case of using the dimensional reduction, although structures similar to integrals of total derivatives were investigated \cite{Aleshin:2015qqc}. At present, the only way to construct the NSVZ scheme with the dimensional reduction is making a specially tuned finite renormalization which relates it with the $\overline{\mbox{DR}}$-scheme in every order of the perturbation theory \cite{Jack:1996vg,Jack:1996cn,Jack:1998uj,Harlander:2006xq,Mihaila:2013wma}. The same situation takes place in theories with softly broken supersymmetry, because renormalization of the softly broken theories is related to renormalization of the rigid theories \cite{Jack:1997eh,Jack:1998iy,Jack:1999aj}. Thus, it is interesting to investigate what is the origin of Eq. (\ref{Soft_Mass_Renormalized}), how it can be directly derived by summing Feynman diagrams, and in what subtraction scheme it is valid. In this paper we give the answers to the first two questions for softly broken ${\cal N}=1$ SQED with $N_f$ flavors, regularized by higher derivatives.

For this purpose we will use the method proposed in \cite{Stepanyantz:2011jy}. Subsequently, it was also applied for obtaining the exact expression for the Adler $D$-function \cite{Adler:1974gd} for ${\cal N}=1$ SQCD in \cite{Shifman:2014cya,Shifman:2015doa}. In this paper we will demonstrate that this method can also be used in softly broken Abelian supersymmetric theories for deriving the exact expression for the photino mass anomalous dimension defined in terms of the bare coupling constant. In particular, we will prove that the renormalization of the photino mass is determined by integrals of double total derivatives in all orders.\footnote{Although the technique proposed in \cite{Avdeev:1997vx} allows to relate the results of calculations in a softly broken theory to the ones in the rigid theory, it would be interesting to derive the exact result by a straightforward calculation.} Exactly as in the case of the rigid theory, such integrals do not vanish due to singularities of the integrands. By summing these singularities for softly broken ${\cal N}=1$ SQED with $N_f$ flavors regularized by higher derivatives it is possible to derive the exact equation

\begin{equation}\label{Exact_Relation}
\frac{d}{d\ln\Lambda}\Big(\frac{m_0}{\alpha_0}\Big) = -\frac{m_0\alpha_0 N_f}{\pi}\cdot \frac{d\gamma(\alpha_0)}{d\alpha_0},
\end{equation}

\noindent
where $m_0$ is the bare photino mass and $\Lambda$ denotes the dimensionful parameter of the regularized theory, which plays the role of the ultraviolet cut-off. (For the considered regularization it is valid in all orders independently of the subtraction scheme.) Differentiating the left hand side and using Eq. (\ref{NSVZ_Beta_Bare}) we obtain the result for the anomalous dimension of the photino mass defined in terms of the bare coupling constant by the prescription

\begin{equation}\label{Gamma_M_Bare}
\gamma_m(\alpha_0) \equiv \frac{d\ln m_0}{d\ln\Lambda},
\end{equation}

\noindent
which is written as

\begin{equation}
\gamma_m(\alpha_0) = \frac{\alpha_0 N_f}{\pi}\Big[1- \frac{d}{d\alpha_0}\Big(\alpha_0\gamma(\alpha_0)\Big)\Big].
\end{equation}

The paper is organized as follows: In Sect. \ref{Section_SQED} we introduce the higher derivative regularization for ${\cal N}=1$ SQED with softly broken supersymmetry.
In the next Sect. \ref{Section_Total_Derivatives} we prove that integrals which determine the renormalization of the photino mass are integrals of double total derivatives by explicit summation of supergraphs in all orders. They are calculated in Sect. \ref{Section_Beta} by summing singularities of the integrands. The result gives Eq. (\ref{Exact_Relation}) in all orders of the perturbation theory. In Sect. \ref{Section_Superfields} we discuss some aspects of renormalization using the $\theta$-dependent renormalization constants. Eq. (\ref{Exact_Relation}) is verified by an explicit two-loop calculation in Sect. \ref{Section_Two_Loop}. In particular, we explicitly construct the integral of a double total derivative which defines renormalization of the photino mass in the two-loop approximation. The results are briefly summarized in the Conclusion. Explicit expressions for various two-loop supergraphs are presented in Appendix.

\section{Softly broken ${\cal N}=1$ SQED and the higher derivative regularization}\label{Section_SQED}
\hspace*{\parindent}

In this paper we consider softly broken $N=1$ SQED with $N_f$ flavours. It is described by the action

\begin{eqnarray}\label{Original_Action}
&& S = \frac{1}{4e_0^2} \mbox{Re} \int d^4x\, d^2\theta\,(1-2m_0\theta^2) W^a W_a + \frac{1}{4} \sum\limits_{f=1}^{N_f} \int d^4x\, d^4\theta\,(1 - \widetilde m_{\phi 0}^2\theta^4) \Big(\phi_f^* e^{2V} \phi_f \qquad\nonumber\\
&& + \widetilde\phi_f^* e^{-2V}\widetilde \phi_f\Big) + \sum\limits_{f=1}^{N_f} \Big(\frac{1}{2}\int d^4x\,d^2\theta\, m_{\phi 0}(1 - \theta^2 b_0)\phi_f \widetilde\phi_f + \mbox{c.c.}\Big),
\end{eqnarray}

\noindent
where $V$ denotes the supersymmetric gauge superfield with the strength $W_a = \bar D^2 D_a V/4$; $\phi_f$ and $\widetilde\phi_f$ are chiral matter superfields. The bare coupling constant is denoted by $e_0$, and $m_{\phi0}$ is the bare mass of the matter superfields in the rigid theory. The soft breaking parameter $m_0$ is the bare photino mass; $\widetilde m_{\phi0}$ and $b_0$ are also the bare soft breaking parameters with the dimension of mass. The soft breaking terms contain the spurion $\eta\equiv \theta^2$. In our notation,\footnote{Below in this paper we prefer to write $\theta$-s instead of $\eta$ and $\bar\eta$.}

\begin{equation}
\theta^2 \equiv \theta^a \theta_a = \eta; \qquad \bar\theta^2 \equiv \bar\theta^{\dot a} \bar\theta_{\dot a} = \bar\eta;\qquad \theta^4 \equiv \theta^2 \bar\theta^2 =\eta\bar\eta.
\end{equation}

\noindent
The considered theory is invariant under $U(1)$ gauge transformations. Due to this symmetry, terms linear and cubic in the matter superfields are forbidden.

In this paper we are interested in the renormalization of the photino mass $m$ in the limit when the parameters $m_\phi$, $\widetilde m_\phi$, and $b$ vanish. This implies that only the following terms in the action will be essential below:

\begin{equation}\label{Action}
S \to \frac{1}{4e_0^2} \mbox{Re} \int d^4x\, d^2\theta\,(1-2m_0\theta^2) W^a W_a + \frac{1}{4} \sum\limits_{f=1}^{N_f} \int d^4x\, d^4\theta\, \Big(\phi_f^* e^{2V} \phi_f + \widetilde\phi_f^* e^{-2V}\widetilde \phi_f\Big).
\end{equation}

\noindent
As for $m$, we will consider only terms linear in $m$ and neglect $m$-dependence of the Green functions. This means that investigating the renormalization of various Green functions we consider the limit in which $m$ is much smaller than the external momenta.

For calculating quantum corrections in the considered theory we will use the higher derivative regularization. It can be introduced by adding the higher derivative term

\begin{equation}\label{Regulator}
S_\Lambda = \frac{1}{4e_0^2} \mbox{Re} \int d^4x\, d^2\theta\,(1-2m_0\theta^2) W^a \Big(R(\partial^2/\Lambda^2)-1\Big) W_a
\end{equation}

\noindent
to the action, where $\Lambda$ is a parameter with the dimension of mass, and $R(x)$ is a regulator rapidly growing at infinity and satisfying the condition $R(0)=1$.  For example, one can choose $R(x)=1+x^n$, where $n$ is an integer. It is convenient to choose the gauge fixing term

\begin{equation}
S_{\mbox{\scriptsize gf}} = -\frac{1}{32 \xi_0 e_0^2} \int d^4x\, d^4\theta\, (1 - m_0\theta^2 - m_0\bar\theta^2)\, D^2V K(\partial^2/\Lambda^2)\bar D^2 V,
\end{equation}

\noindent
where $K$ is a function which rapidly grows at infinity and, by definition, $K(0)=1$. Then the terms quadratic in the gauge superfield can be written as

\begin{eqnarray}\label{Quadratic_Terms}
&& -\frac{1}{4e_0^2} \int d^4x\,d^4\theta\,\Big\{ (1- m_0\theta^2 - m_0 \bar\theta^2) V \Big[R \partial^2\Pi_{1/2}
+ K \frac{1}{16\xi_0}(\bar D^2 D^2 + D^2 \bar D^2) \Big]\, V \nonumber\\
&& + \frac{m_0}{4} V R\cdot \Big[\theta^a \bar D^2 D_a V + \bar\theta^{\dot a} D^2 \bar D_{\dot a}\Big] V - \frac{m_0}{4\xi_0} V K\cdot \Big[\theta^a D_a \bar D^2 + \bar\theta^{\dot a} \bar D_{\dot a} D^2 - \bar D^2 - D^2\Big] V \Big\}.\qquad
\end{eqnarray}

\noindent
Note that the terms in the second string do not contain the second degree of $\theta$-s and come from the derivatives of the spurion. We will calculate quantum corrections by using the supergraph method \cite{Grisaru:1979wc}, which can be also used for theories with softly broken supersymmetry \cite{Yamada:1994id}. We will consider the terms in the first string as the free action. The terms in the second string will be treated as the interaction. They give vertices in which external lines correspond to the first degree $\theta$ polynomials proportional to $m_0$. It is well known that these vertices do not contribute to the considered RG function \cite{Yamada:1994id}. This can also be seen from the calculations made in this paper. According to \cite{HelayelNeto:1984iv,Feruglio:1984aq,Scholl:1984hj} the propagator of the gauge superfield obtained from Eq. (\ref{Quadratic_Terms}) is proportional to\footnote{Constructing this expression one does not take into account the terms in the second string of Eq. (\ref{Quadratic_Terms}), which are considered as interaction.}

\begin{equation}\label{Gauge_Propagator}
e_0^2 (1 + m_0\theta^2 + m_0 \bar \theta^2) \Big[-\frac{1}{R\partial^2} + \frac{1}{16\partial^4}\Big(\frac{\xi_0}{K} - \frac{1}{R}\Big)\Big(\bar D^2 D^2 + D^2 \bar D^2\Big)\Big] \delta^8_{xy} + O(m_0\theta, m_0\bar\theta) + O(m_0^2).
\end{equation}

\noindent
In this equation we omit the terms proportional to $m_0\theta$, $m_0\bar\theta$, and $m_0$ without $\theta$-s (denoted by $O(m_0\theta, m_0\bar\theta)$) and all terms
proportional to $(m_0)^n$ with $n\ge 2$ (denoted by $O(m_0^2)$). Below we will see that they do not affect the considered RG functions. The propagator (\ref{Gauge_Propagator}) contains large degrees of the momentum in the denominator inside the functions $R$ and $K$. Consequently, all diagrams beyond one-loop approximation become finite (at finite values of $\Lambda$). However, divergences can be present in the one-loop graphs \cite{Faddeev:1980be}. According to \cite{Slavnov:1977zf}, these remaining one-loop divergences are regularized by inserting the Pauli--Villars determinants

\begin{eqnarray}\label{PV_Determinants}
&& \mbox{Det}(PV, V, M_I)^{-1} \equiv \int D\Phi_I D\widetilde\Phi_I \exp\Bigg(\frac{i}{4}\int d^4x\,d^4\theta\,\Big(\Phi_I^* e^{2V} \Phi_I + \widetilde\Phi_I^* e^{-2V} \widetilde\Phi_I \Big)\qquad\nonumber\\
&& + \frac{i}{2}\int d^4x\, d^2\theta\, M_I \Phi_I \widetilde\Phi_I + \frac{i}{2}\int d^4x\, d^2\bar\theta\, M_I \Phi_I^* \widetilde\Phi_I^*\Bigg)
\end{eqnarray}

\noindent
into the generating functional. (The superfields $\Phi_I$ and $\widetilde\Phi_I$ are commuting.) Taking into account that in the Abelian case it is not necessary to introduce the Faddeev--Popov ghosts, the generating functional for the connected Green functions can be presented in the form

\begin{equation}\label{W_functional}
W = -i\ln \int DV\, D\phi\, D\widetilde\phi\, \prod\limits_{I=1}^m \mbox{Det}(PV, V, M_I)^{N_f c_I} \exp\Big(i S + i S_\Lambda + i S_{\mbox{\scriptsize gf}} + i S_{\mbox{\scriptsize Sources}}\Big).
\end{equation}

\noindent
The coefficients $c_I$ are introduced in order to cancel the remaining one-loop divergencies. They satisfy the equation

\begin{equation}\label{C_Condition}
\sum_{I=0}^m c_I = 0,
\end{equation}

\noindent
where, by definition, $c_0\equiv -1$. In our conventions the Pauli--Villars masses $M_I$ are proportional to the parameter $\Lambda$ in the higher derivative term $S_\Lambda$,

\begin{equation}\label{M_Lambda_Relation}
M_I = a_I \Lambda,
\end{equation}

\noindent
where the coefficients $a_I$ should not depend on the coupling constant. The action for sources is written as

\begin{equation}
S_{\mbox{\scriptsize Sources}} = \int d^4x\,d^4\theta\, V J + \Big(\int d^4x\,d^2\theta\, (\phi j +\widetilde \phi\, \widetilde j) + \mbox{c.c.}\Big).
\end{equation}

\noindent
The effective action $\Gamma[\bm{V},\phi,\widetilde\phi]$ is obtained from the functional $W$ by making the Legendre transformation. Note that, for later convenience, we denote the argument of the effective action corresponding to the gauge superfield by the bold letter $\bm{V}$. This is done in order to distinguish it from the integration variable $V$ in the generating functional.

The NSVZ-like equation (\ref{Exact_Relation}) relates divergencies in the two-point Green function of the gauge superfield and of the matter superfields. Taking into account the gauge invariance of the action and the gauge invariance of the regularization one can verify that the quantum corrections to the former Green function should be transversal due to the Ward identity. Consequently, the part of the effective action corresponding to this two-point Green function can be presented in the form
\cite{Yamada:1994id}

\begin{eqnarray}\label{Gamma2_V}
&& \Gamma^{(2)}_{\bm{V}} - S_{\mbox{\scriptsize gf}} =  -\frac{1}{16\pi} \int \frac{d^4p}{(2\pi)^4}\,d^4\theta\,\Big(\bm{V}(-p,\theta) \partial^2\Pi_{1/2}\bm{V}(p,\theta)\,  d^{-1}(\alpha_0,\Lambda/p) \nonumber\\
&& - \frac{m_0}{8} \Big(\theta^2 D^a \bm{V}(-p,\theta) \bar D^2 D_a \bm{V}(p,\theta) + \bar\theta^2 \bar D^{\dot a} \bm{V}(-p,\theta) D^2 \bar D_{\dot a} \bm{V}(p,\theta)\Big)\, d_m^{-1}(\alpha_0,\Lambda/p)\Big),\qquad\quad
\end{eqnarray}

\noindent
where, for simplicity we do not write the other arguments of the (dimensionless) functions $d$ and $d_m$, because we are interested only in the behaviour of the function $d_m^{-1}$ in the limit when all massive parameters (except for $\Lambda$ and, therefore, $M_I$) tend to 0. Note that we do not include in this expression a term proportional to ${\displaystyle \int d^4\theta\, \bm{V}}$, because in the considered theory it is forbidden by the $Z_2$-symmetry $V\to -V$; $\phi \leftrightarrow \widetilde\phi$. The normalization constants in Eq. (\ref{Gamma2_V}) are chosen so that in the tree approximation $d^{-1}=\alpha_0^{-1} + O(1)$ and $d_m^{-1} = \alpha_0^{-1} + O(1)$.

The two-point Green function of the chiral matter superfields for theories with softly broken supersymmetry is $\theta$-dependent and can be written as

\begin{equation}\label{Matter_Two_Point_Function_G}
\frac{1}{4}\sum\limits_{f=1}^{N_f}\int \frac{d^4p}{(2\pi)^4}\, d^4\theta\,\left(\phi_f^*(-p,\theta)\, \phi_f(p,\theta) + \widetilde\phi_f^*(-p,\theta)\, \widetilde\phi_f(p,\theta) \right) {\cal G}\left(\alpha_0,\Lambda/p,\theta^2,\bar\theta^2\right),
\end{equation}

\noindent
where

\begin{equation}\label{Matter_Two_Point_Function}
{\cal G}\left(\alpha_0,\Lambda/p,\theta^2,\bar\theta^2\right) = G + m_0 \theta^2 g + m_0 \bar\theta^2 g^* + m_0^2 \theta^4\,\widetilde g.
\end{equation}

\noindent
(The mass $m_0$ is written explicitly in order to make the functions $g$ and $\widetilde g$ dimensionless. This will be convenient below.) The terms linear in the chiral matter superfields are evidently forbidden in the considered theory.

The renormalized coupling constant $\alpha(\alpha_0,\Lambda/\mu)$ and the renormalization constant for the photino mass $Z_m(\alpha,\Lambda/\mu)$, where $\mu$ is a normalization point, are defined by requiring finiteness of the functions

\begin{equation}
d^{-1}(\alpha_0(\alpha,\Lambda/\mu),\Lambda/p)\qquad\mbox{and}\qquad Z_m(\alpha,\Lambda/\mu)\, d_m(\alpha_0(\alpha,\Lambda/\mu),\Lambda/p),
\end{equation}

\noindent
respectively. (The second equation implies that the renormalized photino mass is related to the bare one as $m = Z_m m_0$.)

By definition, the real $\theta$-dependent renormalization constant ${\cal Z}$ is constructed in such a way that the expression

\begin{equation}
{\cal Z}\big(\alpha,\Lambda/\mu,\theta^2,\bar\theta^2 \big)\, {\cal G}\big(\alpha_0,\Lambda/p,\theta^2,\bar\theta^2\big)
\end{equation}

\noindent
is finite. This renormalization constant is obtained from the $\theta$-dependent renormalization of the chiral matter superfields,

\begin{equation}
\phi = {\cal Z}_\phi(\alpha,\Lambda/\mu,\theta^2)\phi_R \equiv \sqrt{Z(\alpha,\Lambda/\mu)}\Big(1 + m_0\theta^2 z(\alpha,\Lambda/\mu)\Big)\phi_R.
\end{equation}

\noindent
(according to this construction the expression $ZG$ is finite) and from the remormalization of the soft scalar masses $\widetilde m^2_\phi$,

\begin{equation}\label{Z_Relation}
{\cal Z} = {\cal Z}_\phi {\cal Z}_\phi^* + O(\theta^4).
\end{equation}

In this paper we investigate the RG functions defined in terms of the bare coupling constant as

\begin{equation}\label{Bare_Definitions}
\beta(\alpha_0) = \frac{d\alpha_0}{d\ln\Lambda}\Big|_{\alpha=\mbox{\scriptsize const}}; \qquad \gamma(\alpha_0) = - \frac{d\ln Z}{d\ln\Lambda}\Big|_{\alpha=\mbox{\scriptsize const}};\qquad \gamma_m(\alpha_0) = - \frac{d\ln Z_m}{d\ln\Lambda}\Big|_{\alpha=\mbox{\scriptsize const}},
\end{equation}

\noindent
where the differentiation with respect to $\ln\Lambda$ should be made at a fixed value of the renormalized coupling constant. (The last equation is evidently equivalent to Eq. (\ref{Gamma_M_Bare}).) As we have already mentioned above, they depend only on the regularization, but are not changed under finite renormalizations (for a fixed regularization).

\section{Integrals of double total derivatives which determine renormalization of the photino mass}\label{Section_Total_Derivatives}

\subsection{Two-point Green function of the gauge superfield}
\hspace*{\parindent}

To obtain the NSVZ-like result for the renormalization of the photino mass in Abelian supersymmetric theories by summing supergraphs, it is possible to use the idea proposed
in \cite{Stepanyantz:2011jy}. It is based on the observation that in this case the action is quadratic in the chiral matter superfields. This implies that the functional
integrals over these matter superfields can formally be calculated exactly in all orders. Following Ref. \cite{Stepanyantz:2011jy}, the result can be presented in the form

\begin{eqnarray}\label{Gaussian_Integral}
&&\int D\phi\,D\widetilde\phi\,\exp\Bigg\{i\Bigg[\frac{1}{4}\int d^4x\,d^4\theta\,\Big(\phi^* e^{2V} \phi + \widetilde\phi^* e^{-2V} \widetilde \phi \Big)
+ \Big(\int d^4x\,d^2\theta\,\Big(\frac{1}{2} M\widetilde\phi\,\phi \qquad\nonumber\\
&& + j\phi + \widetilde j \widetilde \phi\Big) + \mbox{c.c.}\Big)\Bigg]\Bigg\}
= \det (\star)^{1/2}\cdot \exp\Big(\frac{i}{2}\int d^4x\,d^4\theta\, (Aj)^T P \star Aj\Big).\qquad
\end{eqnarray}

\noindent
In this equation the currents are present in the combination

\begin{equation}
Aj \equiv \frac{1}{4\partial^2} \left(\begin{array}{c}D^2 j\\ \bar D^2 j^*\\ D^2\widetilde j\\ \bar D^2 \widetilde j^*\end{array}\right),
\end{equation}

\noindent
corresponding to the sequence $(\phi,\phi^*,\widetilde\phi,\widetilde\phi^*)$ of the matter superfields. The operator

\begin{equation}\label{Star_Definition}
\star \equiv (1-I_0)^{-1},\qquad\mbox{with}\qquad I_0\equiv BP,
\end{equation}

\noindent
encodes chains of the vertices $B$ and the propagators $P$ of arbitrary lengths. The vertices are included into the matrix

\begin{equation}
B \equiv \left(
\begin{array}{cccc}
0 & e^{2V}-1 & 0 & 0\\
e^{2V}-1 & 0 & 0 & 0\\
0 & 0 & 0 & e^{-2V}-1\\
0 & 0 & e^{-2V}-1 & 0
\end{array}
\right),
\end{equation}

\noindent
while the propagator matrix has the form

\begin{equation}
P = \left(
\begin{array}{cccc}
0 & \frac{\bar D^2 D^2}{16(\partial^2 + M^2)} & \frac{M\bar D^2}{4(\partial^2 + M^2)} & 0\\
\frac{D^2\bar D^2}{16(\partial^2 + M^2)} & 0 & 0 & \frac{M D^2}{4(\partial^2 + M^2)} \\
\frac{M \bar D^2}{4(\partial^2 + M^2)} & 0 & 0 & \frac{\bar D^2 D^2}{16 (\partial^2 + M^2)}\\
0 & \frac{MD^2}{4(\partial^2 + M^2)} & \frac{D^2\bar D^2}{16(\partial^2 + M^2)} & 0
\end{array}
\right).
\end{equation}

Using Eq. (\ref{Gaussian_Integral}) the generating functional for the connected Green functions can be presented in the form

\begin{eqnarray}\label{W_Without_j}
&& W = -i \ln \int DV \prod\limits_{I=0}^m \mbox{Det}(PV, V, M_I)^{N_f c_I} \exp\Big(i (S_{\mbox{\scriptsize gauge}} + S_\Lambda + S_{\mbox{\scriptsize gf}}) + i\int d^8x\,J\,V\Big)\qquad
\nonumber\\
&& \times \exp\Big(\frac{i}{2}\int d^8x\,(Aj)^T P \star Aj\Big),
\end{eqnarray}

\noindent
where

\begin{equation}
S_{\mbox{\scriptsize gauge}} = \frac{1}{4e_0^2} \mbox{Re} \int d^4x\, d^2\theta\,(1-2m_0\theta^2) W^a W_a,
\end{equation}

\noindent
and $I=0$ stands for the original superfields $\phi$ and $\widetilde\phi$, so that $M_0=0$ and $c_0 = -1$.

Starting from the expression (\ref{W_Without_j}) and repeating the calculations made in \cite{Stepanyantz:2011jy} we obtain the following expression for the part of the effective action corresponding to the two-point Green function of the gauge superfield:

\begin{eqnarray}\label{General_Form_Of_Gamma}
&& \Delta\Gamma^{(2)}_{\bm V} \equiv \Gamma^{(2)}_{\bm V} -S^{(2)}_{\bm{V}} - S_{\mbox{\scriptsize gf}} = -\frac{i}{2} N_f^2\, \Big\langle \Big(\sum\limits_{I=0}^m c_I \mbox{Tr}(\bm{V} Q J_0\star)\Big)_I^2\Big\rangle_{\mbox{\scriptsize 1PI}}\nonumber\\
&&\qquad\qquad\qquad\qquad\qquad\qquad\ + i N_f \sum\limits_{I=0}^m c_I \mbox{Tr} \Big\langle \bm{V} Q J_0 \star \bm{V} Q J_0 \star + \bm{V}^2 J_0\star \Big\rangle_{I,\mbox{\scriptsize 1PI}},\qquad
\end{eqnarray}

\noindent
where the argument of the effective action is denoted by the bold letter $\bm{V}$ in order to distinguish it from the integration variable $V$. The angular brackets are defined by the prescription

\begin{equation}\label{Angular_Brackets}
\langle A(V) \rangle \equiv \frac{1}{Z} \int DV\, A(V) \prod\limits_{I=0}^m \mbox{Det}(PV, V, M_I)^{N_f c_I} \exp\Big(i (S_{\mbox{\scriptsize gauge}} + S_\Lambda + S_{\mbox{\scriptsize gf}}) + i\int d^8x\,J\,V\Big),
\end{equation}

\noindent
where $Z = \exp(iW)$, and the source $J$ should be expressed in terms of $\bm{V}$ in the standard way by making the Legendre transformation. The subscripts $\mbox{1PI}$ point that it is necessary to take into consideration only one-particle irreducible graphs. Also we use the notation

\begin{equation}
Q \equiv \left(
\begin{array}{cccc}
1 & 0 & 0 & 0\\
0 & 1 & 0 & 0\\
0 & 0 & -1 & 0\\
0 & 0 & 0 & -1
\end{array}
\right);\qquad\qquad J_0 = \left(
\begin{array}{cccc}
0 & e^{2V} & 0 & 0\\
e^{2V} & 0 & 0 & 0\\
0 & 0 & 0 & e^{-2V}\\
0 & 0 & e^{-2V} & 0
\end{array}
\right) P.
\end{equation}

The expression (\ref{General_Form_Of_Gamma}) has a simple graphical interpretation. The first term corresponds to diagrams in which external lines are attached to different
loops of the matter superfields. Usually, it is called the singlet contribution. Below we prove that for the considered theory this contribution vanishes. The second term can be interpreted as a sum of diagrams in which the external $\bm{V}$-lines are attached to different points of a single matter loop. The last term encodes the sum of diagrams
in which external lines are attached to a single point on a matter line.

Although Eq. (\ref{General_Form_Of_Gamma}) looks exactly as a similar expression presented in \cite{Stepanyantz:2011jy} for the rigid theory, there is an essential difference. Namely, the angular brackets are now defined in a different way, because the action $S_{\mbox{\scriptsize gauge}} + S_\Lambda$ contains the soft photino mass term. In particular, this implies that the angular brackets introduce explicit dependence on $\theta$. Therefore, the effective supergraphs which are obtained from Eq. (\ref{W_Without_j}) should be calculated using somewhat different rules in comparison with the rigid theory.

It is convenient to rewrite the expression (\ref{General_Form_Of_Gamma}) in a different form. The corresponding calculation has been done in \cite{Shifman:2015doa}. Here we
only present the result and its graphical interpretation.

\begin{figure}[h]
\begin{picture}(0,6)
\put(4,-0.1){\includegraphics[scale=0.6]{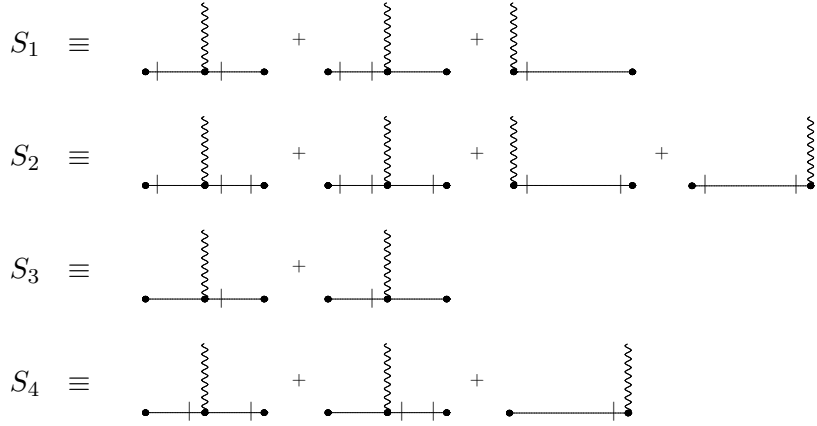}}
\put(2.8,5.15){$S_1\ \ \equiv$}
\put(2.8,3.65){$S_2\ \ \equiv$}
\put(2.8,2.15){$S_3\ \ \equiv$}
\put(2.8,0.65){$S_4\ \ \equiv$}
\end{picture}
\caption{Elements of the matrix $BP (\bm{V}Q) B_0P + B(\bm{V}Q) (\Pi_{+} P) + B(P\Pi_{-})(\bm{V}Q)$ correspond to the sum of subdiagrams presented in this figure.}\label{Figure_V_Subdiagrams}
\end{figure}

The first term in Eq. (\ref{General_Form_Of_Gamma}) contains the expression $\mbox{Tr} (\star \bm{V} Q J_0)$, which can be equivalently presented in the form

\begin{equation}\label{Singlet_Trace}
\mbox{Tr} (\star \bm{V} Q J_0) = \mbox{Tr}\Big\{\star\Big(BP (\bm{V}Q) B_0P + B(\bm{V}Q) (\Pi_{+} P) + B(P\Pi_{-})(\bm{V}Q) \Big)\Big\},
\end{equation}

\noindent
where the chiral projection operators have the form

\begin{equation}
\Pi_{+} \equiv - \frac{\bar D^2 D^2}{16\partial^2};\qquad \Pi_{-} \equiv - \frac{D^2 \bar D^2}{16\partial^2},
\end{equation}

\noindent
and

\begin{equation}
B_0 = \left(
\begin{array}{cccc}
0 & 1 & 0 & 0\\
1 & 0 & 0 & 0\\
0 & 0 & 0 & 1\\
0 & 0 & 1 & 0
\end{array}
\right)
\end{equation}

\noindent
encodes the vertices with a single external $\bm{V}$-line attached to the matter line (and no internal $V$-lines). The sum in the round brackets can be interpreted as a sum of the subdiagrams presented in Fig. \ref{Figure_V_Subdiagrams}, which were first considered in \cite{Stepanyantz:2011jy}. The operator $\star$ gives a sequence of vertices and propagators of an arbitrary length, and $\mbox{Tr}$ converts this chain to a closed loop. The angular brackets in Eq. (\ref{General_Form_Of_Gamma}) make internal gauge lines from all $V$-s in the product of two expressions (\ref{Singlet_Trace}), giving the singlet contribution to the considered two-point function.

According to \cite{Shifman:2015doa} the second term in Eq. (\ref{General_Form_Of_Gamma}) can be presented as a sum of three terms,

\begin{equation}\label{NonSinglet_Contribition}
\mbox{Tr}\langle \star \bm{V} Q J_0 \star \bm{V} Q J_0 \rangle = A_0 + A_1+ A_2
\end{equation}

\begin{figure}[h]
\begin{picture}(0,2.5)
\put(2,0.45){\includegraphics[scale=0.4]{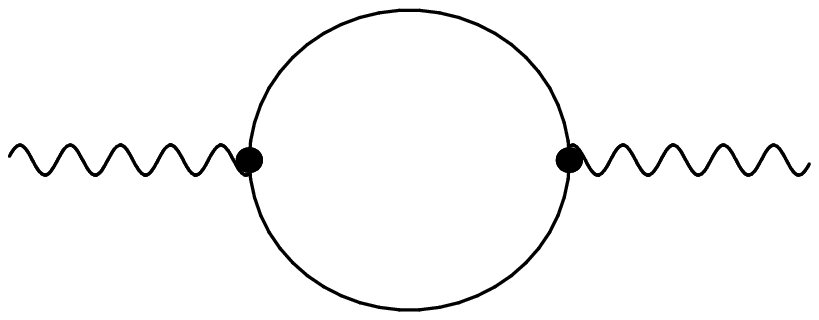}}
\put(6.9,0){\includegraphics[scale=0.4]{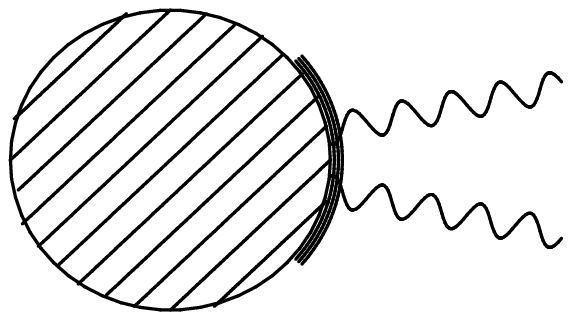}}
\put(12,0.5){\includegraphics[scale=0.4]{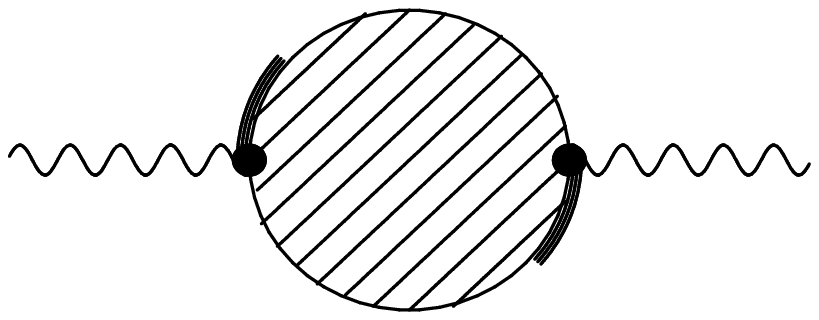}}
\put(0.8,1){$A_0\ = $}
\put(6.2,1){$A_1\ = $}
\put(10.7,1){$A_2\ = $}
\end{picture}
\caption{Graphical interpretation of various terms in Eq. (\ref{NonSinglet_Contribition}).}\label{Figure_Effective_Diagrams}
\end{figure}

\noindent
containing different numbers of the operator $\star$. They can be graphically interpreted as the effective diagrams presented in Fig. \ref{Figure_Effective_Diagrams}. Namely,

\begin{equation}\label{A0}
A_0 \equiv \mbox{Tr} \Big((\bm{V} Q B_0) P (\bm{V} Q B_0) P\Big)
\end{equation}

\noindent
does not contain the operator $\star$ (and, consequently, the angular brackets). It gives the one-loop contribution to the considered two-point function. Therefore, the one-loop approximation in this approach should be considered separately.

\begin{figure}[h]
\hspace*{0.6cm}
\begin{picture}(0,2.5)
\put(0.1,1.6){$U_1\ = $}
\put(1,1){\includegraphics[scale=0.25]{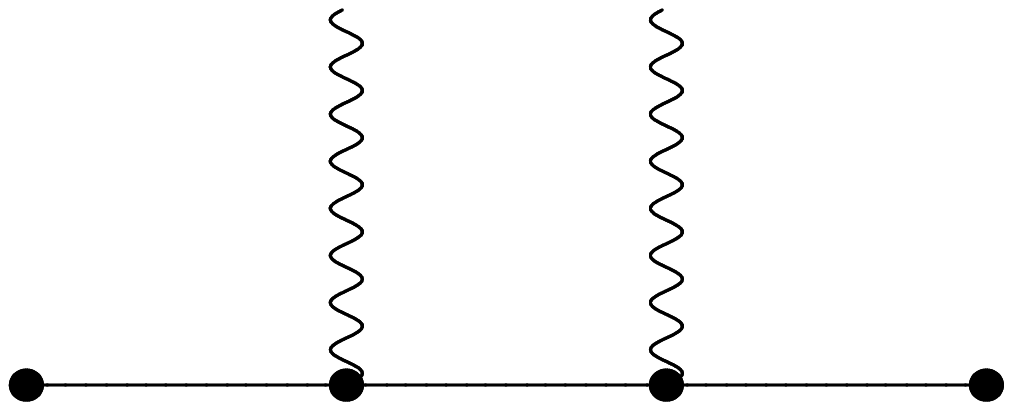}}
\put(1.54,1.24){\mbox{\tiny $|$}} \put(2.36,1.24){\mbox{\tiny $|$}} \put(3.17,1.24){\mbox{\tiny $|$}}
\put(4.25,1.6){$+$}
\put(4.5,1){\includegraphics[scale=0.25]{ver20.eps}}
\put(5.02,1.24){\mbox{\tiny $|$}} \put(5.85,1.24){\mbox{\tiny $|$}} \put(6.35,1.24){\mbox{\tiny $|$}}
\put(7.75,1.6){$+$}
\put(8,1){\includegraphics[scale=0.25]{ver20.eps}}
\put(9.85,1.24){\mbox{\tiny $|$}} \put(8.52,1.24){\mbox{\tiny $|$}} \put(9.05,1.24){\mbox{\tiny $|$}}
\put(11.2,1.6){$+$}
\put(11.5,1){\includegraphics[scale=0.25]{ver20.eps}}
\put(12.02,1.24){\mbox{\tiny $|$}} \put(12.54,1.24){\mbox{\tiny $|$}} \put(13.65,1.24){\mbox{\tiny $|$}}
\put(1.3,0){$+$}
\put(1.7,-0.5){\includegraphics[scale=0.25]{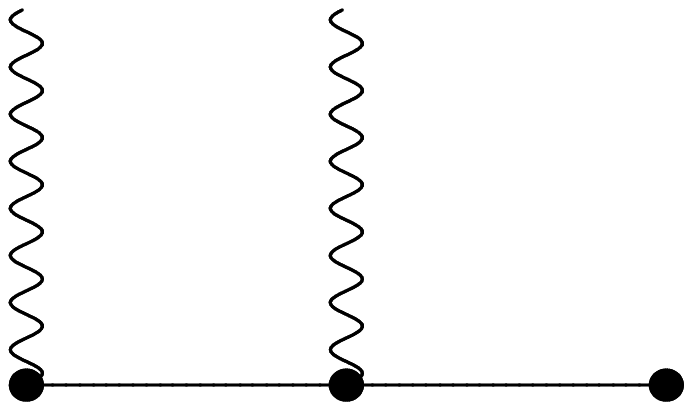}}
\put(2.15,-0.3){\mbox{\tiny $|$}} \put(2.97,-0.3){\mbox{\tiny $|$}}
\put(4.0,0){$+$}
\put(4.4,-0.5){\includegraphics[scale=0.25]{ver22.eps}}
\put(4.85,-0.3){\mbox{\tiny $|$}} \put(5.37,-0.3){\mbox{\tiny $|$}}

\put(0.1,-1.9){$U_2\ = $}
\put(1,-2.5){\includegraphics[scale=0.25]{ver20.eps}}
\put(1.54,-2.26){\mbox{\tiny $|$}} \put(2.36,-2.26){\mbox{\tiny $|$}} \put(3.17,-2.26){\mbox{\tiny $|$}} \put(3.67,-2.26){\mbox{\tiny $|$}}
\put(4.25,-1.9){$+$}
\put(4.5,-2.5){\includegraphics[scale=0.25]{ver20.eps}}
\put(5.02,-2.26){\mbox{\tiny $|$}} \put(5.85,-2.26){\mbox{\tiny $|$}} \put(6.35,-2.26){\mbox{\tiny $|$}} \put(7.18,-2.26){\mbox{\tiny $|$}}
\put(7.75,-1.9){$+$}
\put(8,-2.5){\includegraphics[scale=0.25]{ver20.eps}}
\put(9.85,-2.26){\mbox{\tiny $|$}} \put(8.52,-2.26){\mbox{\tiny $|$}} \put(9.05,-2.26){\mbox{\tiny $|$}} \put(10.68,-2.26){\mbox{\tiny $|$}}
\put(11.2,-1.9){$+$}
\put(11.5,-2.5){\includegraphics[scale=0.25]{ver20.eps}}
\put(12.02,-2.26){\mbox{\tiny $|$}} \put(12.54,-2.26){\mbox{\tiny $|$}} \put(13.65,-2.26){\mbox{\tiny $|$}} \put(14.19,-2.26){\mbox{\tiny $|$}}
\put(1.3,-3.5){$+$}
\put(1.7,-4){\includegraphics[scale=0.25]{ver22.eps}}
\put(2.15,-3.8){\mbox{\tiny $|$}} \put(2.97,-3.8){\mbox{\tiny $|$}} \put(3.5,-3.8){\mbox{\tiny $|$}}
\put(4.0,-3.5){$+$}
\put(4.4,-4){\includegraphics[scale=0.25]{ver22.eps}}
\put(4.85,-3.8){\mbox{\tiny $|$}} \put(5.37,-3.8){\mbox{\tiny $|$}} \put(6.2,-3.8){\mbox{\tiny $|$}}
\put(6.75,-3.5){$+$}
\put(7.0,-4.04){\includegraphics[scale=0.25]{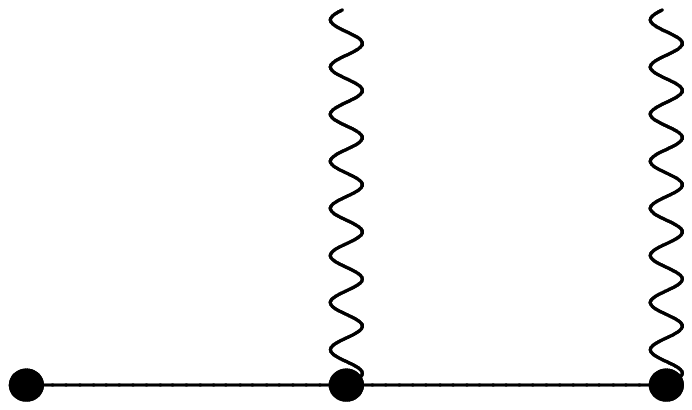}}
\put(7.5,-3.8){\mbox{\tiny $|$}} \put(8.04,-3.8){\mbox{\tiny $|$}} \put(8.85,-3.8){\mbox{\tiny $|$}}
\put(9.45,-3.5){$+$}
\put(9.7,-4.04){\includegraphics[scale=0.25]{ver21.eps}}
\put(10.2,-3.8){\mbox{\tiny $|$}} \put(11.02,-3.8){\mbox{\tiny $|$}} \put(11.55,-3.8){\mbox{\tiny $|$}}
\put(12.08,-3.5){$+$}
\put(12.3,-4.01){\includegraphics[scale=0.25]{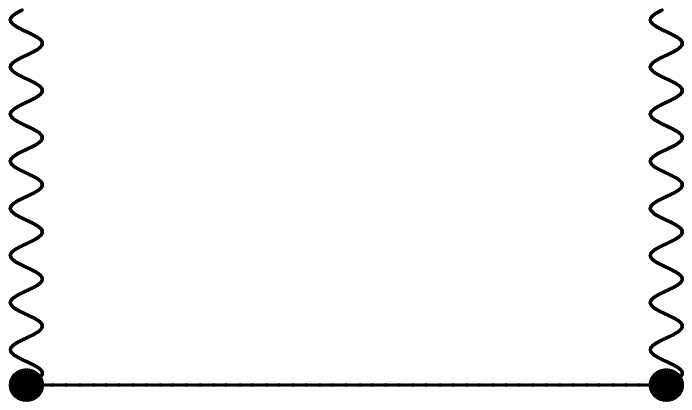}}
\put(12.85,-3.8){\mbox{\tiny $|$}} \put(14.2,-3.8){\mbox{\tiny $|$}}

\put(0.1,-5.4){$U_3\ = $}
\put(1,-6){\includegraphics[scale=0.25]{ver20.eps}}
\put(2.04,-5.76){\mbox{\tiny $|$}} \put(2.86,-5.76){\mbox{\tiny $|$}}
\put(4.25,-5.4){$+$}
\put(4.5,-6){\includegraphics[scale=0.25]{ver20.eps}}
\put(5.55,-5.76){\mbox{\tiny $|$}} \put(6.65,-5.76){\mbox{\tiny $|$}}
\put(7.75,-5.4){$+$}
\put(8,-6){\includegraphics[scale=0.25]{ver20.eps}}
\put(9.34,-5.76){\mbox{\tiny $|$}} \put(10.15,-5.76){\mbox{\tiny $|$}}
\put(11.2,-5.4){$+$}
\put(11.5,-6){\includegraphics[scale=0.25]{ver20.eps}}
\put(12.83,-5.76){\mbox{\tiny $|$}} \put(13.37,-5.76){\mbox{\tiny $|$}}

\put(0.1,-7.4){$U_4\ = $}
\put(1,-8){\includegraphics[scale=0.25]{ver20.eps}}
\put(2.04,-7.76){\mbox{\tiny $|$}} \put(2.86,-7.76){\mbox{\tiny $|$}} \put(3.67,-7.76){\mbox{\tiny $|$}}
\put(4.25,-7.4){$+$}
\put(4.5,-8){\includegraphics[scale=0.25]{ver20.eps}}
\put(5.55,-7.76){\mbox{\tiny $|$}} \put(6.65,-7.76){\mbox{\tiny $|$}} \put(7.18,-7.76){\mbox{\tiny $|$}}
\put(7.75,-7.4){$+$}
\put(8,-8){\includegraphics[scale=0.25]{ver20.eps}}
\put(9.32,-7.76){\mbox{\tiny $|$}} \put(10.15,-7.76){\mbox{\tiny $|$}} \put(10.7,-7.75){\mbox{\tiny $|$}}
\put(11.2,-7.4){$+$}
\put(11.5,-8){\includegraphics[scale=0.25]{ver20.eps}}
\put(12.82,-7.76){\mbox{\tiny $|$}} \put(13.39,-7.76){\mbox{\tiny $|$}} \put(14.20,-7.76){\mbox{\tiny $|$}}
\put(1.3,-9){$+$}
\put(1.6,-9.55){\includegraphics[scale=0.25]{ver21.eps}}
\put(2.64,-9.3){\mbox{\tiny $|$}} \put(3.45,-9.3){\mbox{\tiny $|$}}
\put(4.0,-9){$+$}
\put(4.3,-9.55){\includegraphics[scale=0.25]{ver21.eps}}
\put(5.6,-9.3){\mbox{\tiny $|$}} \put(6.15,-9.3){\mbox{\tiny $|$}}
\end{picture}

\vspace*{9.5cm}

\caption{Subdiagrams corresponding to the effective vertex in the effective diagram $A_1$ presented in Fig. \ref{Figure_Effective_Diagrams}. Note that various options of the
end chiralities lead to four different sums of subdiagrams.}\label{Figure_Double_Subdiagrams}
\end{figure}

The expression

\begin{eqnarray}\label{A1}
&& A_1 = 2\cdot \mbox{Tr}\Big\langle \star \Big(BP (\bm{V}QB_0) P (\bm{V}QB_0) P + (\bm{V} B Q) (\Pi_+ P) (\bm{V} Q B_0) P
\nonumber\\
&& + BP(\bm{V}QB_0) (P\Pi_{-})(\bm{V}Q) + (B\bm{V}Q)(\Pi_{+} P \Pi_{-}) (\bm{V}Q)\Big)\Big\rangle
\end{eqnarray}

\noindent
is constructed from the terms containing a single operator $\star$. Graphically, it can be presented as the second effective diagram in Fig. \ref{Figure_Effective_Diagrams}. The effective vertex in this diagram consists of a large number of subdiagrams with two external $\bm{V}$-lines. They are presented in Fig. \ref{Figure_Double_Subdiagrams}.

The last term in Eq. (\ref{NonSinglet_Contribition}) is a sum of contributions containing two operators $\star$. It is explicitly written as

\begin{eqnarray}\label{A2}
&& A_2 = \mbox{Tr}\,\Big\langle \star \Big(BP(\bm{V}Q)B_0P + B(\bm{V}Q)(\Pi_{+}P) + B(P\Pi_{-})(\bm{V}Q)\Big)\nonumber\\
&& \star \Big(BP(\bm{V}Q)B_0P + B(\bm{V}Q)(\Pi_{+}P) + B(P\Pi_{-})(\bm{V}Q)\Big) \Big\rangle
\end{eqnarray}

\noindent
and can be graphically presented as the third diagram in Fig. \ref{Figure_Effective_Diagrams}. Each of two effective vertices in this diagram is given by the sum of
subdiagrams presented in Fig. \ref{Figure_V_Subdiagrams}. As usual, the operators $\star$ and $\mbox{Tr}$ make the closed loop, and the angular brackets transform $V$ into
gauge propagators.

Below we will see that the last term in Eq. (\ref{General_Form_Of_Gamma}) (which contains $\bm{V}^2$) is not essential in calculating the considered RG function, because it gives the vanishing contribution. That is why we do not discuss in details its structure.

To calculate the left hand side of Eq. (\ref{Exact_Relation}), it is convenient to consider the expression

\begin{equation}\label{We_Calculate}
\frac{d}{d\ln\Lambda}\Big(\frac{m_0}{d_m(\alpha_0,\Lambda/p)} -\frac{m_0}{\alpha_0}\Big)\Big|_{p=0} = -\frac{d}{d\ln\Lambda}\Big(\frac{m_0}{\alpha_0}\Big),
\end{equation}

\noindent
where the differentiation is made at fixed values of the renormalized coupling constant and of the renormalized photino mass. The equality follows from the finiteness of the expression $m_0 d_m^{-1} = m (Z_m d_m)^{-1}(\alpha,\mu/p,\Lambda/p)$. The limit $p\to 0$ should be taken in order to get rid of the various terms proportional to $(p/\Lambda)^k$.

The left hand side of Eq. (\ref{We_Calculate}) can be obtained from the part of $\Delta\Gamma = \Gamma -S- S_{\mbox{\scriptsize gf}}$ corresponding to the two-point Green function of the gauge superfield. We denote it by $\Delta\Gamma^{(2)}_{\bm{V}}$. By other words, $\Delta\Gamma^{(2)}_{\bm{V}}$ is a part of $\Delta\Gamma$ which contains only terms quadratic in the gauge superfield $\bm{V}$. To extract the expression (\ref{We_Calculate}), it is possible to make the formal substitution

\begin{equation}\label{Substitution_For_V}
\bm{V} \to \bar\theta^{\dot a} \bar\theta_{\dot a} \theta^b \psi_b \equiv \bar\theta^2 \theta^b \psi_b,
\end{equation}

\noindent
where $\psi_b$ is a slowly changing spinor which is approximately constant at finite $x^\mu$ and tends to 0 at some large scale $R\to\infty$. Really, after this substitution
the part of the gauge superfield two-point function corresponding to the rigid theory (which contains the function $d^{-1}$) vanishes, because the supersymmetric transversal
projection operator $\partial^2 \Pi_{1/2}$ contains four spinor derivatives. Note that the derivatives $\partial_\mu$ acting on $\psi$ give terms suppressed as $1/R\Lambda$,
because $\psi$ is almost constant. These terms are negligible in the limit $R\to\infty$ and should be omitted. However, the part of the gauge superfield two-point function
containing the function $d_m^{-1}$ survives after the substitution (\ref{Substitution_For_V}) and, after some transformations, gives

\begin{equation}\label{We_Calculate_Gamma}
\frac{d\Delta\Gamma^{(2)}_{\bm{V}}}{d\ln\Lambda}\Big|_{\bm{V} = \bar\theta^2 \theta^b \psi_b} = - \frac{1}{8\pi} {\cal V}_\psi\cdot \frac{d}{d\ln\Lambda}\Big(\frac{m_0}{d_m(\alpha_0,\Lambda/p)} -\frac{m_0}{\alpha_0}\Big)\Big|_{p=0},
\end{equation}

\noindent
where we introduced the notation

\begin{equation}\label{V_Psi}
{\cal V}_\psi \equiv \int d^4 x\,\psi^a \psi_a \sim R^4 \to \infty.
\end{equation}

\noindent
(Note that the condition $p\to 0$ appears, because the spinor $\psi_b$ is almost constant. It follows from the fact that the external momentum $p$ has the order $1/R$
according to the definition of the spinor $\psi$.) It is important that the only non-vanishing (after the substitution (\ref{Substitution_For_V})) term in Eq. (\ref{Gamma2_V}) is the one containing $m_0 \theta^2$, while the term containing $m_0\bar\theta^2$ vanishes. That is why it is also possible to set formally

\begin{equation}\label{Theta_Substitution}
m_0\bar\theta^2 \to 0
\end{equation}

\noindent
in $S_{\mbox{\scriptsize gauge}}$ and $S_\Lambda$ before the substitution (\ref{Substitution_For_V}), as we will always do below. (However, $m_0 \theta^2\ne 0$.)

\subsection{Factorization into double total derivatives}
\hspace*{\parindent}

Let us calculate the expression (\ref{We_Calculate_Gamma}) by using Eq. (\ref{General_Form_Of_Gamma}). As we have demonstrated in the previous section, the result can be
obtained by calculating some effective diagrams which contain certain subdiagrams. That is why we start with the calculation of the subdiagrams presented in Fig.
\ref{Figure_V_Subdiagrams}, in which we make the substitution (\ref{Substitution_For_V}). The analytical expressions for them are encoded in the expression

\begin{eqnarray}\label{Analytical_Form_Of_Subdiagrams}
&& BP(\bm{V}Q)B_0P + B(\bm{V}Q)(\Pi_{+}P) + B(P\Pi_{-})(\bm{V}Q)\vphantom{\Big(}\nonumber\\
&&\qquad = \left(
\begin{array}{cccc}
(e^{2V}-1) S_4 & 0 & 0 & (e^{2V}-1) S_3\\
0 & (e^{2V}-1) S_1 & (e^{2V}-1) S_2 & 0\\
0 & - (e^{-2V}-1) S_3 & - (e^{-2V}-1) S_4 & 0\\
- (e^{-2V}-1) S_2 & 0 & 0 & -(e^{-2V}-1) S_1
\end{array}
\right),\qquad
\end{eqnarray}

\noindent
where $S_1$, $S_2$, $S_3$, and $S_4$ denote expressions for the subdiagrams presented in Fig. \ref{Figure_V_Subdiagrams}. Omitting the derivatives of the spinor $\psi$, which vanish in the limit $p\to 0$, after some transformations the result can be written as

\begin{eqnarray}\label{Expressions_For_Subdiagrams}
&& S_1 = - \bar\theta^{\dot a} \theta^b \psi_b \frac{\bar D_{\dot a} D^2}{4(\partial^2 + M^2)} + i\bar\theta^{\dot a} (\gamma^\mu)_{\dot a}{}^b \psi_b \frac{\bar D^2 D^2 \partial_\mu}{16(\partial^2 + M^2)^2} + \mbox{terms without $\bar\theta$}\nonumber\\
&&\qquad\qquad\qquad\qquad\qquad\qquad\qquad\, = \frac{i}{2} \bar\theta \gamma^\mu \psi \Big[y_\mu^*, \frac{\bar D^2 D^2}{16(\partial^2+M^2)}\Big] + \mbox{terms without $\bar\theta$};\qquad\nonumber\\
&& S_2 = -\bar\theta^{\dot a} \theta^b \psi_b \frac{M \bar D_{\dot a}}{\partial^2 + M^2} + i \bar\theta^{\dot a} (\gamma^\mu)_{\dot a}{}^b \psi_b \frac{M \bar D^2 \partial^\mu}{4(\partial^2 + M^2)^2} + \mbox{terms without $\bar\theta$}\nonumber\\
&&\qquad\qquad\qquad\qquad\qquad\qquad\qquad\ \ = \frac{i}{2} \bar\theta \gamma^\mu \psi \Big[y_\mu^*,\frac{M\bar D^2}{4(\partial^2 + M^2)}\Big] + \mbox{terms without $\bar\theta$};\nonumber\\
&& S_3 = - i\bar\theta^{\dot a} (\gamma^\mu)_{\dot a}{}^b \psi_b \frac{M D^2 \partial_\mu}{4(\partial^2 + M^2)^2} + \mbox{terms without $\bar\theta$}\nonumber\\
&&\qquad\qquad\qquad\qquad\qquad\qquad\qquad\,  = -\frac{i}{2}\bar\theta \gamma^\mu \psi \Big[y_\mu^*, \frac{M D^2}{4(\partial^2 + M^2)}\Big] + \mbox{terms without $\bar\theta$};\nonumber\\
&& S_4 = -i \bar\theta^{\dot a} (\gamma^\mu)_{\dot a}{}^b \psi_b \frac{D^2 \bar D^2 \partial_\mu}{16 (\partial^2 + M^2)^2} + \bar\theta^{\dot a} \theta^b \psi_b \frac{D^2 \bar D_{\dot a}}{4(\partial^2 + M^2)} + \bar\theta^{\dot a} \psi^b \frac{D_b \bar D_{\dot a}}{2(\partial^2 + M^2)} \nonumber\\
&&\qquad\qquad\qquad + \mbox{terms without $\bar\theta$} = -\frac{i}{2} \bar\theta \gamma^\mu \psi \Big[y_\mu^*, \frac{D^2 \bar D^2}{16(\partial^2 + M^2)}\Big] + \mbox{terms without $\bar\theta$},
\end{eqnarray}

\noindent
where $\bar\theta \gamma^\mu \psi \equiv \bar\theta^{\dot a} (\gamma^\mu)_{\dot a}{}^b \psi_b$. $M$ is equal either to 0 for the usual matter superfields, or to $M_I$ for the Pauli--Villars superfields. The (anti)chiral coordinates are defined by

\begin{equation}
y^\mu = x^\mu + i \bar\theta^{\dot a} (\gamma^\mu)_{\dot a}{}^b \theta_b;\qquad (y^\mu)^* = x^\mu - i \bar\theta^{\dot a} (\gamma^\mu)_{\dot a}{}^b \theta_b.
\end{equation}

\noindent
Below we will see that the terms without $\bar\theta$ do not contribute to the considered RG function (which determines the running photino mass). Looking at the expressions
(\ref{Expressions_For_Subdiagrams}) we see that all terms in the right hand side contain commutators of $y_\mu^*$ with various components of the matrix $P$. Therefore, from Eqs. (\ref{Analytical_Form_Of_Subdiagrams}) and (\ref{Expressions_For_Subdiagrams}) we obtain

\begin{equation}\label{Sum_Of_Subdiagrams}
BP(\bm{V}Q)B_0P + B(\bm{V}Q)(\Pi_{+}P) + B(P\Pi_{-})(\bm{V}Q) = \frac{i}{2} \bar\theta \gamma^\mu \psi [y_\mu^*,\widetilde Q I_0] + \mbox{terms without $\bar\theta$},
\end{equation}

\noindent
where

\begin{equation}
\widetilde Q \equiv \left(
\begin{array}{cccc}
-1 & 0 & 0 & 0\\
0 & 1 & 0 & 0\\
0 & 0 & 1 & 0\\
0 & 0 & 0 & -1
\end{array}
\right).
\end{equation}

\noindent
(It is easy to see that $[\widetilde Q,I_0] = 0$ and $[\widetilde Q, \star]=0$.) Then we make the transformations similar to the ones in \cite{Stepanyantz:2011jy,Shifman:2015doa}, taking into account that trace of the commutator with $-i\bar\theta^{\dot a} (\gamma^\mu)_{\dot a}{}^b \theta_b$ vanishes, so that $y_\mu^*$ can be replaced by $x_\mu$. As a result, we obtain

\begin{equation}\label{Single_Loop}
\mbox{Tr} (\bm{V} Q J_0 \star)\Big|_{\bm{V}=\bar\theta^2 \theta^a \psi_a} = \frac{i}{2} \mbox{Tr}\Big(\bar\theta^{\dot a} (\gamma^\mu)_{\dot a}{}^b \psi_b \widetilde Q [x_\mu, \ln \star]\Big) + \ldots,
\end{equation}

\noindent
where dots denote terms which do not contain $\bar\theta$. The commutator with $x^\mu$ in the momentum representation gives the derivative with respect to the loop momentum, and $\mbox{Tr}$ gives the integration over this loop momentum. Therefore, we obtain that loop integrals encoded in Eq. (\ref{Single_Loop}) are integrals of total derivatives.

According to Eq. (\ref{General_Form_Of_Gamma}), the sum of diagrams in which the external lines are attached to different matter loops contains two traces (\ref{Single_Loop}). The terms without $\bar\theta$ do not contribute to such diagrams, because the nontrivial result can be obtained only if $\theta^4$ is present in a supergraph. (Let us remind that we have set $\bar\theta^2 =0$ in $S_{\mbox{\scriptsize gauge}}$ and $S_\Lambda$.) Thus, the considered part of the effective action proportional to $N_f^2$ can be presented in the form of an integral of a double total derivative

\begin{equation}\label{Singlet_Contribution}
\frac{i}{8} N_f^2 \Big\langle\Big(\mbox{Tr} \sum\limits_{I=0}^m c_I \bar \theta^{\dot a} (\gamma^\mu)_{\dot a}{}^b \psi_b\, \widetilde Q \left[x_\mu, \ln (\star_I)\right]\Big)^2 \Big\rangle.
\end{equation}

\noindent
More exactly, each matter loop to which an external line is attached gives an integral over a total derivative. Evidently, all diagrams in the considered contribution contain two such loops. It is important that there are no singularities in Eq. (\ref{Singlet_Contribution}), which can produce $\delta$-functions. (We will discuss such singularities in detail below.)

The main difference of Eq. (\ref{Singlet_Contribution}) from the corresponding expressions in Refs. \cite{Stepanyantz:2011jy,Shifman:2015doa} is that it contains only
$\bar\theta^2$ instead of $\theta^4$ in the rigid theory. In the rigid theory all contribution which contain $\bar\theta^2$ and do not contain $\theta^2$ vanish. Really, explicit $\theta$-dependence cannot appear in calculating supergraphs by using the algebra of supersymmetric covariant derivatives. Therefore, any non-vanishing
superdiagram should contain $\theta^4$ due to the integration over $d^4\theta$.

However, the situation is different in the softly broken theory, because the gauge propagators (\ref{Gauge_Propagator}) and vertices coming from the second string of Eq.
(\ref{Quadratic_Terms}) explicitly depend on $\theta$. Therefore, $\theta^2$ is introduced by the functional integration over the gauge superfield which is denoted by the
angular brackets. Evidently, all terms containing $\theta^2$ are proportional at least to the first degree of $m_0$. This implies that the expression
(\ref{Singlet_Contribution}) does not vanish due to the integration over $d^4\theta$. However, it certainly vanishes as a trace of a commutator.

Now, let us proceed to calculating the terms proportional to $N_f$ in Eq. (\ref{General_Form_Of_Gamma}). First, we note that the last term proportional to $\bm{V}^2$ vanishes after the substitution (\ref{Substitution_For_V}) due to anticommutation of $\bar\theta$-components ($\bar\theta^2 \cdot \bar\theta^2 = 0$). Certainly, this result is quite reasonable, because this term is not transversal. Such terms cancel each other due to the Ward identity and do not affect the RG functions considered in this paper.

Thus, there is the only non-vanishing term in Eq. (\ref{General_Form_Of_Gamma}) which is proportional to the expression (\ref{NonSinglet_Contribition}). As we saw above, it can be written as a sum of three contributions $A_0$, $A_1$, and $A_2$, which are graphically presented in Fig. \ref{Figure_Effective_Diagrams}. It is easy to see that the one-loop contribution $A_0$ vanishes after the substitution (\ref{Substitution_For_V}),

\begin{equation}
A_0\Big|_{\bm{V} = \bar\theta^2 \theta^a \psi_a} = \mbox{One-loop} = 0.
\end{equation}

\noindent
Really, the expression $A_0$ is determined by the diagram without internal lines of the gauge superfield. Therefore, $m_0$ does not enter in this supergraph. All other massive parameters (except for the regularization parameter $\Lambda$) are set to 0. Therefore, this diagram is exactly the same as in the rigid theory and contains contributions proportional to $\int d^4\theta \bm{V} \partial \Pi_{1/2} \bm{V}$ and to $\int d^4\theta \bm{V}^2$. It is easy to see that both these expressions vanish after the substitution (\ref{Substitution_For_V}).

The expression $A_1$ is graphically presented in Fig. \ref{Figure_Effective_Diagrams}, where it corresponds to the second effective diagram. The effective vertex in this diagram is given by the expression in the round brackets in Eq. (\ref{A1}). This expression encodes the sums of vertices presented in Fig. \ref{Figure_Double_Subdiagrams}:

\begin{eqnarray}\label{Analytical_Form_Of_A1}
&&\hspace*{-5mm} BP (\bm{V}QB_0) P (\bm{V}QB_0) P + (\bm{V} B Q) (\Pi_+ P) (\bm{V} Q B_0) P
+ BP(\bm{V}QB_0) (P\Pi_{-})(\bm{V}Q)  + (B\bm{V}Q) \vphantom{\Big(}\nonumber\\
&&\hspace*{-5mm} \times (\Pi_{+} P \Pi_{-}) (\bm{V}Q) = \left(
\begin{array}{cccc}
(e^{2V}-1) U_4 & 0 & 0 & (e^{2V}-1) U_3\\
0 & (e^{2V}-1) U_1 & (e^{2V}-1) U_2 & 0\\
0 & (e^{-2V}-1) U_3 & (e^{-2V}-1) U_4 & 0\\
(e^{-2V}-1) U_2 & 0 & 0 & (e^{-2V}-1) U_1
\end{array}
\right).\qquad
\end{eqnarray}

\noindent
Note that after the substitution (\ref{Substitution_For_V}) the external lines correspond to $\bar\theta^2 \theta^a \psi_a$. The sums of subdiagrams presented in Fig. \ref{Figure_Double_Subdiagrams} are given by the following expressions:

\begin{eqnarray}
&&\hspace*{-3mm} U_1 = \frac{1}{16} \bar\theta^2 \psi^2 \Big(-\theta^2 \frac{2 D^2}{\partial^2 + M^2} + \theta^a \frac{i(\gamma^\mu)_{a}{}^{\dot b}\bar D_{\dot b} D^2 \partial_\mu}{(\partial^2 + M^2)^2} + \frac{M^2 \bar D^2 D^2}{2(\partial^2 + M^2)^3}\Big)+\mbox{terms without $\bar\theta^2$}\nonumber\\
&&\hspace*{-3mm} = -\frac{1}{16} \bar\theta^2 \psi^2 \Big[(y^\mu)^*,\Big[y_\mu^*,\frac{\bar D^2 D^2}{16(\partial^2 + M^2)}\Big]\Big]+\mbox{terms without $\bar\theta^2$};\nonumber\\
&&\hspace*{-3mm} U_2 = \frac{1}{16} \bar\theta^2 \psi^2 \Big(-\theta^2 \frac{8 M}{\partial^2 + M^2}+ \frac{4iM}{(\partial^2 + M^2)^2} \theta^{a} (\gamma^\mu)_{a}{}^{\dot b} \bar D_{\dot b}\partial_\mu + \frac{2M^3}{(\partial^2 + M^2)^3} \bar D^2\Big)\nonumber\\
&&\hspace*{-3mm} +\mbox{terms without $\bar\theta^2$} = -\frac{1}{16} \bar\theta^2 \psi^2 \Big[(y^\mu)^*, \Big[y_\mu^*, \frac{M \bar D^2}{4(\partial^2 + M^2)}\Big]\Big]+\mbox{terms without $\bar\theta^2$};\nonumber\\
&&\hspace*{-3mm} U_3 = \bar\theta^2 \psi^2 \frac{M^3}{8(\partial^2 + M^2)^3} D^2 = -\frac{1}{16} \bar\theta^2 \psi^2 \Big[(y^\mu)^*, \Big[y_\mu^*, \frac{M D^2}{4(\partial^2 + M^2)}\Big]\Big]; \vphantom{\frac{1}{2}}\nonumber\\
&&\hspace*{-3mm} U_4 = \frac{1}{16} \bar\theta^2 \psi^2 \Big(-\frac{2}{\partial^2 + M^2} D^2 \theta^2 - \frac{i (\gamma^\mu)_{\dot a}{}^b}{(\partial^2 + M^2)^2}  D^2 \bar D^{\dot a} \theta_b \partial_\mu + \frac{M^2}{2(\partial^2 + M^2)^3} D^2 \bar D^2\Big)\nonumber\\
&&\hspace*{-3mm} +\mbox{terms without $\bar\theta^2$} = -\frac{1}{16} \bar\theta^2 \psi^2 \Big[(y^\mu)^*, \Big[y_\mu^*, \frac{D^2 \bar D^2}{16(\partial^2 + M^2)}\Big]\Big]+\mbox{terms without $\bar\theta^2$}.
\end{eqnarray}

\noindent
(Note that formally writing the equalities we omit possible singular contributions in the massless case, which will be discussed later in details.) Thus, in the matrix form the result can be formally presented as

\begin{eqnarray}
&& BP (\bm{V}QB_0) P (\bm{V}QB_0) P + (\bm{V} B Q) (\Pi_+ P) (\bm{V} Q B_0) P + BP(\bm{V}QB_0) (P\Pi_{-})(\bm{V}Q)\vphantom{\Big(}\qquad\nonumber\\
&& + (B\bm{V}Q)(\Pi_{+} P \Pi_{-}) (\bm{V}Q) = -\frac{1}{16} \bar\theta^2 \psi^2 \left[ (y^\mu)^*, \left[y_\mu^*, I_0\right]\right] + \mbox{terms without $\bar\theta^2$}.\vphantom{\Big)}
\end{eqnarray}

\noindent
Let us substitute this expression into Eq. (\ref{A1}). Then the terms without $\bar\theta^2$ vanish. Really, the non-trivial result is obtained only if a
supergraph contains $\theta^4$, but the functional integration can produce only $\theta$-s (and cannot produce $\bar\theta$-s, see Eq. (\ref{Theta_Substitution})). Therefore,

\begin{equation}
A_1 = -\frac{1}{8} \mbox{Tr}\, \Big\langle \bar\theta^2 \psi^2 \star \left[(y^\mu)^*,\left[y_\mu^*, I_0\right]\right]\Big\rangle.
\end{equation}

The expression $A_2$ is given by the last diagram in Fig. \ref{Figure_Effective_Diagrams}. It contains two effective vertices. Each of these effective vertices can be presented as a sum of subdiagrams presented in Fig. \ref{Figure_V_Subdiagrams}. As we discussed above, after the substitution (\ref{Substitution_For_V}) they can be written in the form (\ref{Sum_Of_Subdiagrams}). Substituting two these expressions into Eq. (\ref{A2}) and taking into account vanishing of the terms which do not contain $\bar\theta^2$, we obtain

\begin{equation}\label{A2_Final_Expression}
A_2 = -\frac{1}{8} \mbox{Tr}\,\Big\langle \bar\theta^2 \psi^2 \left[(y^\mu)^*, I_0\right] \star \left[y_\mu^*,I_0\right]\star \Big\rangle.
\end{equation}

After some transformations similar to the ones described in \cite{Shifman:2015doa}, the sum of the contributions $A_0$, $A_1$, and $A_2$ can be presented in the form

\begin{equation}\label{A_Sum}
A_0 + A_1 + A_2 = -\frac{1}{8} \mbox{Tr}\,\Big\langle \bar\theta^2 \psi^2 \left[(y^\mu)^*,\left[y_\mu^*,\ln(\star)\right]\right]\Big\rangle -\mbox{singularities}.
\end{equation}

\noindent
Note that deriving this equation it is necessary to commute $\bar\theta^2$ with $\star$ and $I_0$. Such commutators are no more than linear in $\bar\theta$ and, therefore, disappear after integration over the anticommuting variables.

The traces of commutators in Eq. (\ref{A_Sum}) evidently vanish, but the nontrivial result is obtained due to singularities, as we explain below. Note that traces of $\theta$ commutators do not produce the singularities, so that in Eq. (\ref{A_Sum}) we can write $x^\mu$ instead of $(y^\mu)^*$. Therefore, the final result for the considered expression can be presented in the form

\begin{eqnarray}\label{Result}
&& \frac{d\Delta\Gamma^{(2)}_{\bm{V}}}{d\ln\Lambda}\Big|_{\bm{V} = \bar\theta^2 \theta^b \psi_b} = \frac{i}{8} N_f^2 \Big\langle\Big(\mbox{Tr} \sum\limits_{I=0}^m c_I \bar \theta^{\dot a} (\gamma^\mu)_{\dot a}{}^b \psi_b\, \widetilde Q \left[x_\mu, \ln (\star_I)\right]\Big)^2 \Big\rangle\nonumber\\
&& - \frac{iN_f}{8} \frac{d}{d\ln\Lambda} \sum\limits_{I=0}^m c_I \mbox{Tr} \Big\langle \bar\theta^2 \psi^2 \left[x^\mu,\left[x_\mu,\ln(\star)\right]\right]\Big\rangle_I -\mbox{singularities} + O(m_0^2).\qquad
\end{eqnarray}

\noindent
We see that in the momentum representation the right hand side is given by integrals of double total derivatives, because the commutator with $x^\mu$ corresponds to the
derivative with respect to the momentum of the loop to which the external lines are attached. The trace gives integration over this momentum. However, the integrals of total
derivatives do not vanish, because the integrands contain singularities. Really, if $f$ is a function, which rapidly tends to 0 at infinity, and $q^\mu$ is the Euclidean
momentum, then

\begin{eqnarray}\label{Momentum_Integral}
&& \int \frac{d^4q}{(2\pi)^4} \frac{\partial}{\partial q^\mu}\Big(\frac{q^\mu}{q^4} f(q^2) \Big) = 2\pi^2 \int \frac{dq^2}{(2\pi)^4} \frac{d}{d q^2} f(q^2) = -\frac{1}{8\pi^2} f(0)\nonumber\\
&&\qquad\qquad\qquad\qquad\qquad = -2\pi^2 \int \frac{d^4q}{(2\pi)^4} f(q^2) \delta^4(q) = - \int \frac{d^4q}{(2\pi)^4} \frac{\partial}{\partial q^\mu}\Big(\frac{q^\mu}{q^4}\Big) f(q^2).\qquad
\end{eqnarray}

\noindent
Note that in the left hand side of this equation, by definition, we can commute $\partial/\partial q^\mu$ and $q^\mu/q^4$, because a small vicinity of $q=0$ is not included into the integration domain. Now, let us define the operator $\bm{\partial}/\bm{\partial} q^\mu$ which, by definition, does not commute with $q^\mu/q^4$ and satisfies the relation

\begin{equation}
\int \frac{d^4q}{(2\pi)^4} \frac{\bm{\partial}}{\bm{\partial} q^\mu} A^\mu = \mbox{Tr} [x_\mu, A^\mu] = 0.
\end{equation}

\noindent
Then Eq. (\ref{Momentum_Integral}) can be rewritten in the form

\begin{equation}
\int \frac{d^4q}{(2\pi)^4} \frac{q^\mu}{q^4} \frac{\partial}{\partial q^\mu} f(q^2) =
\int \frac{d^4q}{(2\pi)^4} \frac{\bm{\partial}}{\bm{\partial} q^\mu}\Big(\frac{q^\mu}{q^4} f(q^2) \Big)
- \int \frac{d^4q}{(2\pi)^4} \frac{\partial}{\partial q^\mu}\Big(\frac{q^\mu}{q^4}\Big) f(q^2).
\end{equation}

\noindent
The first term in the right hand side corresponds to the trace of the commutator in Eq. (\ref{Result}) and vanishes, and the second one comes from the singularity.

\section{Exact expression for the photino mass RG function}\label{Section_Beta}
\hspace*{\parindent}

In the previous section we obtained that the loop integrals which determine the renormalization of the photino mass are integrals of double total derivatives in the limit of the vanishing external momentum. However, these integrals do not vanish, because the integrands contain singularities proportional to $1/q^2$, where $q$ is the Euclidean momentum. In this section we find the sum of singular contributions and compare it with the two-point function of the matter superfields.

First, we note that contributions of the massive Pauli--Villars superfields cannot contain singularities proportional to $1/q^2$. Therefore, only the terms corresponding to
$I=0$ (for which $c_0=-1$ and $M_0=0$) in Eq. (\ref{Result}) do not vanish. Singularities cannot also appear in the singlet contribution, because there are only the first derivatives of $1/q^2$ in it. Therefore, singularities can arise only in the non-singlet contribution for $I=0$. In this case

\begin{equation}
\star = \left(
\begin{array}{cccc}
\bar{*} & 0 & 0 & 0\\
0 & * & 0 & 0\\
0 & 0 & \bar{\tilde{*}} & 0\\
0 & 0 & 0 & \tilde{*}
\end{array}
\right),
\end{equation}

\noindent
where

\begin{eqnarray}
&& * \equiv \frac{1}{1-(e^{2V}-1)\bar D^2 D^2/16\partial^2};\qquad\ \ \bar{*} \equiv \frac{1}{1-(e^{2V}-1) D^2 \bar D^2/16\partial^2};\nonumber\\
&& \tilde{*} \equiv \frac{1}{1-(e^{-2V}-1)\bar D^2 D^2/16\partial^2};\qquad\, \bar{\tilde{*}} \equiv \frac{1}{1-(e^{-2V}-1) D^2 \bar D^2/16\partial^2}.
\end{eqnarray}

\noindent
By making the substitution $V \to - V$ in the generating functional, it is easy to see that the contributions containing $*$ and $\tilde{*}$ ($\bar{*}$ and $\bar{\tilde{*}}$) are equal. Moreover, it is possible to verify that the contributions of $*$ and $\bar{*}$ are also equal. To see this, one should note that they can be related by reversing the sequence of the operators $D$, $\bar D$ etc. Therefore, we obtain

\begin{equation}
\frac{d\Delta\Gamma^{(2)}_{\bm{V}}}{d\ln\Lambda}\Big|_{\bm{V} = \bar\theta^2 \theta^a \psi_a} = \frac{i N_f}{2} \frac{d}{d\ln\Lambda} \mbox{Tr}\, \psi^2 \Big\langle \bar\theta^2 [(y_\mu)^*, [(y^\mu)^*, \ln(*)]]\Big\rangle - \mbox{singularities}.
\end{equation}

\noindent
The left hand side of this equation is related to the function $d_m^{-1}(\alpha_0,\Lambda/p)$ by Eq. (\ref{We_Calculate_Gamma}). To calculate the expression in the right hand side, we start with calculating the inner commutator. Note that, due to the operator $\mbox{Tr}$, cyclic permutations in the expression which is commuted with the first
$(y_\mu)^*$ do not change singular contributions. This follows from the fact that $\bar\theta^2$ in the above expression can be shifted to an arbitrary point of the
considered supergraph (because the terms containing $\bar\theta$ in lower degrees evidently vanish.) Taking into account the possibility of making these cyclic permutations,
we can write the result of calculating the inner commutator in the form

\begin{eqnarray}\label{Singularities}
&& - \frac{1}{8\pi} {\cal V}_\psi \cdot \frac{d}{d\ln\Lambda}\Big(\frac{m_0}{d_m(\alpha_0,\Lambda/p)} - \frac{m_0}{\alpha_0}\Big)\Big|_{p=0} = N_f \frac{d}{d\ln\Lambda} \mbox{Tr}\, \psi^2 \Big\langle \bar\theta^2 \Big[(y_\mu)^*,
i (e^{2V}-1) \frac{\bar D^2 D^2 \partial_\mu}{16\partial^4} * \nonumber\\
&& - (\gamma^\mu)^{c\dot d} (e^{2V}-1) \theta_c \frac{\bar D_{\dot d} D^2}{8\partial^2} * \Big]\Big\rangle - \mbox{singularities}.\vphantom{\frac{1}{2}}
\end{eqnarray}

The trace of the commutator is evidently equal to 0, but the result does not vanish due to singularities, which can appear both from the first term and from the second term. To calculate the singularity of the first term, we use the identity

\begin{equation}\label{Singularity_Origin}
\Big[(y^\mu)^*, \frac{\partial_\mu}{\partial^4} \Big] = \Big[-i\frac{\partial}{\partial q_\mu},-\frac{iq_\mu}{q^4}\Big] = -2\pi^2 \delta^4(q_E) = -2\pi^2 i \delta^4(q).
\end{equation}

\noindent
Then the contribution of the first term in Eq. (\ref{Singularities}) can be presented as

\begin{eqnarray}\label{First_Term}
&& \frac{\pi^2}{8} N_f \frac{d}{d\ln\Lambda} \mbox{Tr}\, \psi^2 \Big\langle \bar\theta^2 * (e^{2V}-1) \bar D^2 D^2 \delta^4(\partial)\Big\rangle\nonumber\\
&&\qquad\qquad\qquad\qquad = -\frac{\pi^2}{128} N_f \frac{d}{d\ln\Lambda} \mbox{Tr}\, \psi^2 \Big\langle \bar\theta^2 * (e^{2V}-1) \frac{\bar D^2 D^2}{\partial^2} \delta^4(\partial) \bar D^2 D^2 \Big\rangle.\qquad
\end{eqnarray}

\noindent
Terms proportional to the first degree of $m_0$ in this expression can be expressed via the two-point Green function of the matter superfields. To do this, we
note that the operator $\mbox{Tr}$ contains the integration over $d^4\theta$. The non-trivial result can be obtained only if the integrand contains $\theta^4 = \bar\theta^2 \cdot \theta^2$. Therefore, the covariant derivatives cannot act to $\bar\theta^2$ explicitly written in Eq. (\ref{First_Term}), and the expression (\ref{First_Term}) can be presented in the form

\begin{equation}
-\frac{\pi^2}{128} N_f \frac{d}{d\ln\Lambda} \mbox{Tr}\, \psi^2 \Big\langle \bar\theta^2 \bar D^2 D^2 * (e^{2V}-1) \frac{\bar D^2 D^2}{\partial^2} \delta^4(\partial) \Big\rangle.
\end{equation}

\noindent
(As usual, we omit all terms with the derivatives of the slowly varying spinor $\psi$, which are suppressed by powers of $\Lambda^{-1}$.) In the momentum representation the argument of the $\delta$-function becomes a loop momentum. In particular,

\begin{equation}
\delta^4(\partial) \delta^4(x-y) = \int \frac{d^4q}{(2\pi)^4} \delta^4(q) e^{-iq_\alpha (x^\alpha-y^\alpha)} = \frac{1}{(2\pi)^4}.
\end{equation}

\noindent
Using this identity we present the first singular contribution in Eq. (\ref{Singularities}) in the form

\begin{equation}
- \frac{N_f}{32\pi^2} \frac{d}{d\ln\Lambda} \int d^4x\,d^4y\,d^4\theta_x\,\psi_x^2\, \bar\theta_x^2\, \Big\langle \frac{\bar D_x^2 D_x^2}{8} * \frac{\bar D_x^2 D_x^2}{8\partial^2}\delta^8_{xy} \Big\rangle\Big|_{\theta_y = \theta_x}.
\end{equation}

\noindent
From the other hand, differentiating Eq. (\ref{Gaussian_Integral}) with respect to the sources $j_x$ and $j_y^*$ for each flavor we obtain

\begin{eqnarray}\label{Matter_Green_Function}
\Big(\frac{\delta^2\Gamma}{\delta\phi_y^* \delta\phi_x}\Big)^{-1}= - \frac{\delta^2 W}{\delta j_y^* \delta j_x} = \Big\langle \frac{\bar D_x^2 D_x^2}{8\partial^2} * \frac{\bar D_x^2 D_x^2}{8\partial^2} \delta^8_{xy} \Big\rangle,
\end{eqnarray}

\noindent
where (in the limit, when all masses except for $m_0$ vanish) the inverse Green function satisfies the condition

\begin{equation}\label{Inverse_Function_Definition}
\int d^8y\, \frac{\delta^2\Gamma}{\delta\phi_x \delta\phi_y^*} \frac{\bar D_y^2}{8\partial^2} \Big(\frac{\delta^2\Gamma}{\delta\phi_z \delta\phi_y^*}\Big)^{-1} = -\frac{1}{2} \bar D_x^2 \delta^8_{xz}.
\end{equation}

\noindent
Therefore, it is possible to relate the considered expression to the two-point Green function of the matter superfields by the equation

\begin{eqnarray}\label{First_Singularity_Vs_Two_Point}
- \frac{N_f}{32\pi^2} \frac{d}{d\ln\Lambda} \int d^4x\,d^4y\,d^4\theta_x\,\psi_x^2\, \bar\theta_x^2\, \partial_x^2 \Big(\frac{\delta^2\Gamma}{\delta\phi_x \delta\phi_y^*}\Big)^{-1}\Big|_{\theta_y=\theta_x}.
\end{eqnarray}

The two-point function of the matter superfields can be easily found by differentiating Eq. (\ref{Matter_Two_Point_Function}),

\begin{equation}
\frac{\delta^2\Gamma}{\delta\phi_x \delta\phi_y^*} = \frac{1}{16} D_y^2 \Big(G + m_0\theta^2 g + m_0 \bar\theta^2 g^* + m_0^2\theta^4 \widetilde g\Big)_y  \bar D_x^2 \delta^8_{xy}.
\end{equation}

\noindent
The corresponding inverse function can be found from Eq. (\ref{Inverse_Function_Definition}). Taking into account that due to Eq. (\ref{Theta_Substitution}) the dependence on $\bar\theta$ disappears, the result can be written as

\begin{equation}
\Big(\frac{\delta^2\Gamma}{\delta\phi_x \delta\phi_y^*}\Big)^{-1} \to - \frac{\bar D_x^2 D_y^2}{4\partial^2} (G + m_0\theta^2 g)^{-1}\delta^8_{xy}.
\end{equation}

\noindent
Substituting the expression for the inverse Green function to Eq. (\ref{First_Singularity_Vs_Two_Point}) we obtain

\begin{equation}
\frac{N_f}{128\pi^2} \frac{d}{d\ln\Lambda} \int d^4x\,d^4y\,d^4\theta_x\,\psi_x^2\, \bar\theta_x^2 \bar D_x^2 D_y^2 \left(G + m_0\theta^2 g\right)^{-1}\delta^8_{xy} \Big|_{\theta_y=\theta_x}.
\end{equation}

\noindent
To simplify this expression, we note that the covariant derivatives do not act to the explicitly written $\theta$-s, because the integral over $d^4\theta$ does not vanish only if the integrand contains $\theta^4$. Therefore, they should act to $\delta^8_{xy}$,

\begin{equation}
\bar D_x^2 D_y^2 \delta^8_{xy} \Big|_{\theta_y=\theta_x} = 4 \delta^4(x-y).
\end{equation}

\noindent
The coordinate $\delta$-function allows calculating one of the coordinate integrals. The integrand in the remaining coordinate integral contains the spinor $\psi$, which slowly depends on the coordinates. In the momentum representation this implies that the corresponding momentum tends to 0. Therefore, using Eq. (\ref{V_Psi}) we obtain that the first term of Eq. (\ref{Singularities}) gives

\begin{equation}\label{Result_For_The_First_Term}
{\cal V}_\psi \cdot N_f \frac{1}{32\pi^2} \frac{d}{d\ln\Lambda} \int d^4\theta\,\bar\theta^2 \left(G + m_0\theta^2 g\right)^{-1}\Big|_{q=0} + O(m_0^2),
\end{equation}

\noindent
where the momentum $q$ is an argument of the functions $G$ and $g$. Note that this expression is not well-defined. The well-defined expression will be obtained after adding the contribution of the second term in Eq. (\ref{Singularities}).

\begin{figure}[h]
\begin{picture}(0,2)
\put(4,0){\includegraphics[scale=0.22]{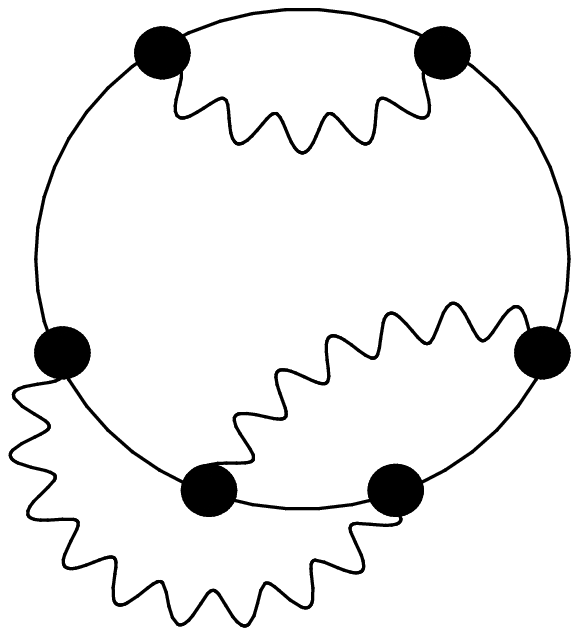}}
\put(9,0.3){\includegraphics[scale=0.068]{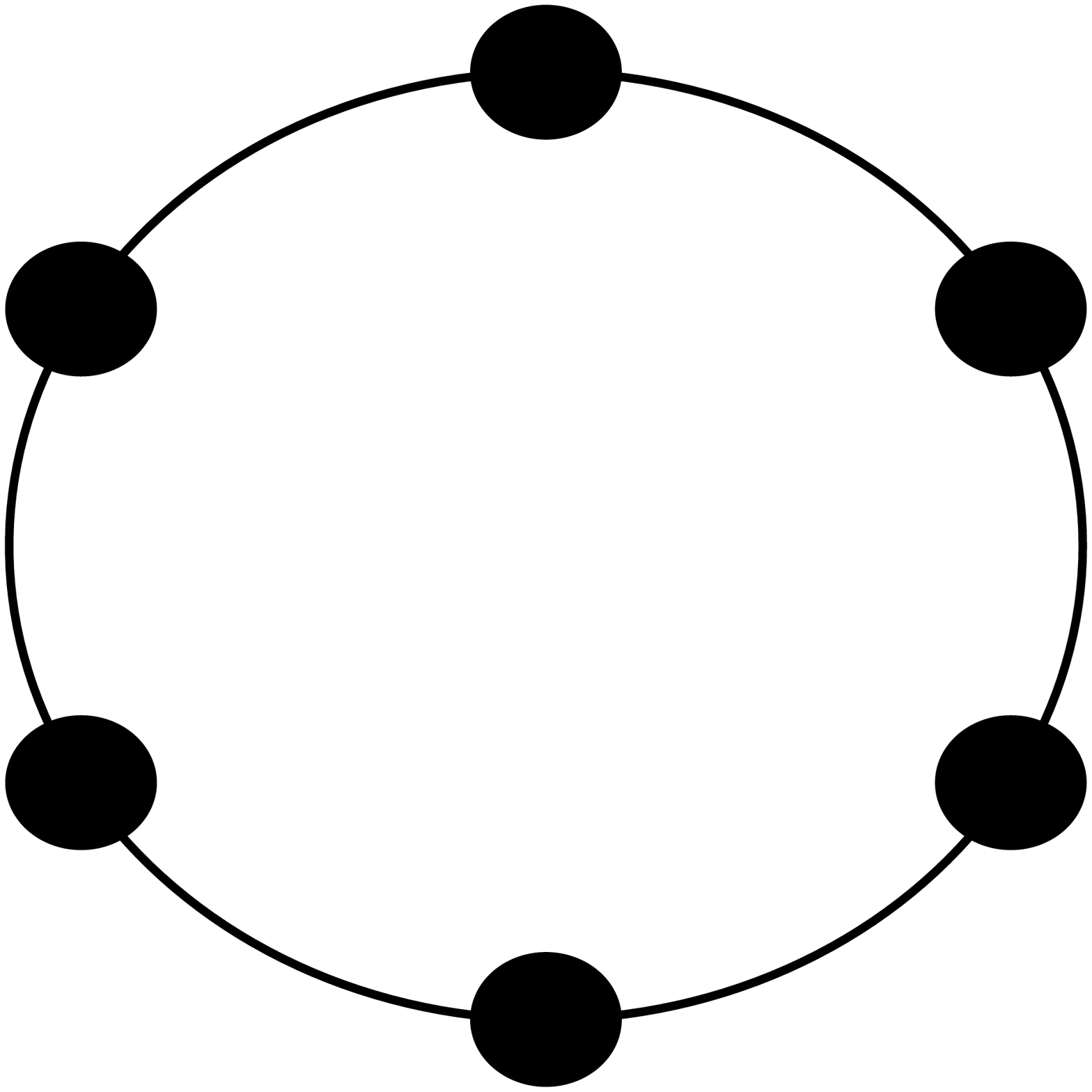}}
\end{picture}
\label{Figure_Double_Singularities}
\caption{Supergraphs which have coinciding momenta. The left supergraph corresponds to $n=2$, and the right one schematically illustrates the $n=6$ case. In the right diagram the circles denote the 1PI subdiagrams contributing to the anomalous dimension of the matter superfields.}
\end{figure}

Singularities coming from the second term appear only if a diagram has coinciding momenta in some propagators of the matter loop. An example of such a diagram is given in the left part of Fig. \ref{Figure_Double_Singularities}. From this figure, it is evident that coinciding momenta appear only if the corresponding graph can be made disconnected by two cuts of the matter line. By other words, it consists of some 1PI parts which are connected with each other only by the matter line, see the right part of Fig. \ref{Figure_Double_Singularities}, where these parts are denoted by the circles. Let us denote a number of these parts by $n$.

We have already mentioned that due to integration over $d^4\theta$ the integrand should contain $\theta^4$. $\bar\theta^2$ is explicitly present in Eq. (\ref{Singularities}). $\theta^2$ (or $\theta$) is contained inside the gauge propagators (\ref{Gauge_Propagator}) and vertices with smaller degrees of $\theta$-s. In the first term of Eq. (\ref{Singularities}) $\theta^2$ can appear from each of the $n$ 1PI parts of the considered supergraph. In this term the covariant derivatives do not act to $\theta^2$, because then the degree of $\theta$ will be less than 4. Therefore, the expression for the first term contains

\begin{equation}\label{First_Term_Contribution}
m_0 \theta^2 \Big(g_1 \Delta G_2 \ldots \Delta G_n + \Delta G_1 g_2 \ldots \Delta G_n + \ldots + \Delta G_1 \Delta G_2 \ldots g_n \Big),
\end{equation}

\noindent
where $G \equiv 1+\Delta G$ and $\Delta G_i$ and $g_i$ are contributions of 1PI parts to the functions $G$ and $g$, respectively.

Analysing the second term of Eq. (\ref{Singularities}), first, we consider the case when the supersymmetric covariant derivatives do not act to $\theta^2$. In this case we obtain $(n-1)$ contributions containing

\begin{equation}
-\frac{\bar D^2 D^2}{16\partial^2}\cdot (\gamma^\mu)^{c\dot d} \theta_c \frac{\bar D_{\dot d} D^2}{4\partial^2} = \frac{i\partial^\mu \bar D^2 D^2}{8\partial^4}
\end{equation}

\noindent
and 1 contribution proportional to $\theta^2 \theta_c = 0$. Therefore, (after symmetrization) the sum of the considered terms will be proportional to

\begin{equation}\label{Second_Term_Contribution}
- \frac{(n-1)}{n} m_0 \theta^2 \Big(g_1 \Delta G_2 \ldots \Delta G_n + \Delta G_1 g_2 \ldots \Delta G_n + \ldots +  \Delta G_1 \Delta G_2 \ldots g_n \Big).
\end{equation}

\noindent
Also it is necessary to consider the case when the supersymmetric covariant derivatives act to $\theta^2$. Because the second term of Eq. (\ref{Singularities}) is linear
in the explicitly written $\theta$, only one $D$ can act to $\theta^2$. However, from the dimensional and chirality considerations it is possible to see that in this case the
supergraph contains the structures like

\begin{equation}
\theta_a \cdot D^2 \cdot (\gamma^\mu)^{c\dot d} \theta_c \frac{\bar D_{\dot d} D^2}{16\partial^2} = O(\theta)\qquad \mbox{or} \qquad
(\gamma^\nu)^{a\dot b} \theta_a \bar D_{\dot b} \cdot (\gamma^\mu)^{c\dot d} \theta_c \frac{\bar D_{\dot d} D^2}{16\partial^2} = \theta^2 \frac{\bar D^2 D^2}{32\partial^2},
\end{equation}

\noindent
where $O(\theta)$ denotes terms linear in $\theta$ and terms without $\theta$. In the first case the corresponding contributions vanish, and there are no
singularities (coming from $\partial_\mu/\partial^4$) in the second case.

\begin{figure}[h]
\begin{picture}(0,4)
\put(2.5,0.9){\includegraphics[scale=0.3]{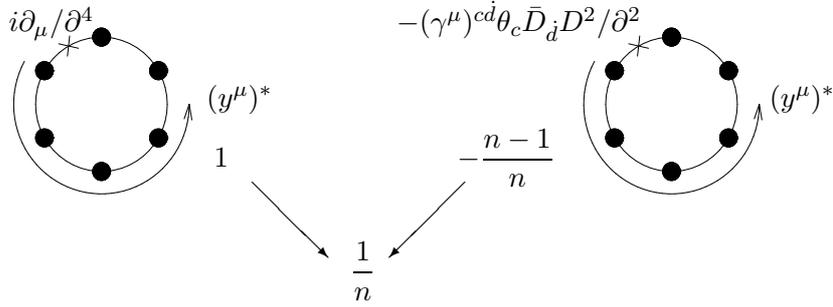}}
\put(10.0,0.9){\includegraphics[scale=0.3]{businki1.eps}}
\put(5.5,1.5){$1$}
\put(8.7,1.5){${\displaystyle - \frac{n-1}{n}}$}
\put(6,1.3){\vector(1,-1){1}}
\put(8.8,1.3){\vector(-1,-1){1}}
\put(7.3,0){${\displaystyle \frac{1}{n}}$}
\put(5.4,2.3){$(y^\mu)^*$}
\put(12.8,2.3){$(y^\mu)^*$}
\put(2.8,3.3){$i\partial_\mu/\partial^4$}
\put(7.9,3.3){$-(\gamma^\mu)^{c\dot d}\theta_c \bar D_{\dot d} D^2/\partial^2$}
\end{picture}
\label{Figure_Singularities}
\caption{Summation of singularities.}
\end{figure}

Consequently, comparing Eqs. (\ref{First_Term_Contribution}) and (\ref{Second_Term_Contribution}), we see that the contribution of the second term in Eq. (\ref{Singularities}) is equal to the contribution of the first term multiplied by $(1-n)/n$. This implies that the sum of both terms is in

\begin{equation}
1 - \frac{n-1}{n} = \frac{1}{n}
\end{equation}

\noindent
times larger than the contribution of the first term. The expression for first term (given by Eq. (\ref{Result_For_The_First_Term})) contains

\begin{equation}
(G + m_0\theta^2 g)^{-1} = (1 + \Delta G + m_0\theta^2 g)^{-1} \to m_0 \theta^2 \sum\limits_{n=1}^\infty n (-1)^n g (\Delta G)^{n-1},
\end{equation}

\noindent
where the arrow points that we omit terms without $\theta^2$, which give vanishing contributions. Therefore, the sum of both terms is proportional to

\begin{equation}
m_0 \theta^2 \sum\limits_{n=1}^\infty (-1)^n g (\Delta G)^{n-1} = - m_0 \theta^2 g (1 + \Delta G)^{-1} = - m_0 \theta^2 g G^{-1}.
\end{equation}

\noindent
Thus, in order to find the sum of all singular terms, it is necessary to replace $(G+m_0 \theta^2 g)^{-1}$ in Eq. (\ref{Result_For_The_First_Term}) by
$-m_0 \theta^2 G^{-1} g$. The integration over the anticommuting variable $\theta$ gives the multiplier 4. Therefore, if we denote the momentum by $q$, the all-orders result
for the expression (\ref{We_Calculate_Gamma}) takes the form

\begin{equation}\label{All_Loop_Result}
- {\cal V}_\psi \cdot \frac{N_f}{8\pi^2} \frac{d}{d\ln\Lambda} \Big(m_0 G(\alpha_0,\Lambda/q)^{-1} g(\alpha_0,\Lambda/q)\Big)\Big|_{q=0} + O(m_0^2).\qquad
\end{equation}

\noindent
Note that in Eq. (\ref{We_Calculate_Gamma}) the differentiation with respect to $\ln\Lambda$ is made at fixed values of the renormalized coupling constant and the renormalized photino mass. The first term in the right hand side of Eq. (\ref{We_Calculate_Gamma}) is a finite function of these parameters. This implies that its derivative with respect to $\ln\Lambda$ vanishes. Therefore, comparing Eqs. (\ref{We_Calculate_Gamma}) and (\ref{All_Loop_Result}) in the limit $m_0 \to 0$, we obtain the equation which does not contain the auxiliary parameter ${\cal V}_\psi \to \infty$,

\begin{equation}\label{RG_Equation}
\frac{d}{d\ln\Lambda}\Big(\frac{m_0}{\alpha_0}\Big) = - \frac{N_f}{\pi} \frac{d}{d\ln\Lambda}\Big(m_0 G(\alpha_0,\Lambda/q)^{-1} g(\alpha_0,\Lambda/q)\Big)\Big|_{q=0}.
\end{equation}

\noindent
The dependence of supergraphs on $m_0$ and $\alpha_0$ comes from the propagator of the gauge superfield (\ref{Gauge_Propagator}) and the vertices with smaller degrees of
$\theta$-s. The function $g(\alpha_0,\Lambda/q)$ is determined by terms proportional to $m_0 \theta^2$. Therefore, the vertices in the second string of Eq.
(\ref{Quadratic_Terms}) and terms proportional to $m_0$ without $\theta^2$ in the gauge propagators (\ref{Gauge_Propagator}) do not contribute to this function. Consequently, from Eq. (\ref{Gauge_Propagator}) it is evident that the $m_0\theta^2$ terms can be obtained by differentiating the superdiagram for the rigid theory with respect to $\ln\alpha_0$, namely,

\begin{equation}\label{Expression_For_g}
g(\alpha_0,\Lambda/q) = \alpha_0 \frac{\partial}{\partial\alpha_0}  G(\alpha_0,\Lambda/q).
\end{equation}

\noindent
This implies that the right hand side of Eq. (\ref{RG_Equation}) can be presented in the form

\begin{eqnarray}\label{RG_Function}
&& -\frac{N_f}{\pi} \frac{d}{d\ln\Lambda} \Big(m_0 \alpha_0 \frac{\partial}{\partial \alpha_0} \ln G(\alpha_0,\Lambda/q)\Big)\Big|_{q=0} = -\frac{N_f}{\pi} \frac{m_0}{\alpha_0} \frac{d}{d\ln\Lambda} \Big(\alpha_0^2 \frac{\partial}{\partial \alpha_0} \ln G(\alpha_0,\Lambda/q)\Big)\Big|_{q=0}\qquad\nonumber\\
&& -\frac{N_f}{\pi} \frac{d}{d\ln\Lambda}\Big(\frac{m_0}{\alpha_0} \Big) \alpha_0^2 \frac{\partial}{\partial \alpha_0} \ln G(\alpha_0,\Lambda/q)\Big|_{q=0}.\qquad
\end{eqnarray}

\noindent
In this expression the total derivative $d/d\ln\Lambda$ acts to both explicitly written $\ln\Lambda$ and $\ln\Lambda$ inside $\alpha_0$. It can be expressed in terms of the partial derivatives as

\begin{equation}\label{Total_Derivative}
\frac{d}{d\ln\Lambda} = \beta(\alpha_0) \frac{\partial}{\partial \alpha_0} + \frac{\partial}{\partial\ln\Lambda},
\end{equation}

\noindent
where the partial derivative $\partial/\partial\ln\Lambda$ acts only to the explicitly written $\ln\Lambda$. By the help of Eq. (\ref{Total_Derivative}) one can commute the derivatives entering Eq. (\ref{RG_Function}), taking into account that

\begin{equation}
\gamma(\alpha_0) = -\frac{d\ln Z}{d\ln\Lambda} = \frac{d\ln G(\alpha_0,\Lambda/q)}{d\ln\Lambda}\Big|_{q=0};\qquad \beta(\alpha_0) = \frac{d\alpha_0}{d\ln\Lambda}.
\end{equation}

\noindent
As the result, we rewrite the expression (\ref{RG_Function}) in the form

\begin{eqnarray}\label{After_Commuting}
&& \frac{d}{d\ln\Lambda}\Big(\frac{m_0}{\alpha_0}\Big) = -\frac{N_f}{\pi} \Big[ m_0\alpha_0 \frac{d\gamma(\alpha_0)}{d\alpha_0} - m_0\alpha_0^3 \frac{d}{d\alpha_0}\Big(\frac{\beta(\alpha_0)}{\alpha_0^2}\Big) \frac{\partial}{\partial\alpha_0}\ln G(\alpha_0,\Lambda/q)\nonumber\\
&& + \frac{d}{d\ln\Lambda}\Big(\frac{m_0}{\alpha_0} \Big) \alpha_0^2 \frac{\partial}{\partial \alpha_0} \ln G(\alpha_0,\Lambda/q) \Big|_{q=0}\Big].\qquad\quad
\end{eqnarray}

\noindent
With the higher derivative regularization the $\beta$-function and the anomalous dimension of the matter superfields defined in terms of the bare coupling constant satisfy the NSVZ relation in all orders (in an arbitrary subtraction scheme). This was proved in \cite{Stepanyantz:2011jy,Stepanyantz:2014ima} by direct summation of supergraphs. Therefore, it is possible to use Eq. (\ref{NSVZ_Beta_Bare}), so that

\begin{equation}
\frac{d}{d\alpha_0}\Big(\frac{\beta(\alpha_0)}{\alpha_0^2}\Big) = -\frac{N_f}{\pi} \frac{d\gamma(\alpha_0)}{d\alpha_0}.
\end{equation}

\noindent
We substitute this expression into Eq. (\ref{After_Commuting}). This gives

\begin{eqnarray}
&&\ \frac{d}{d\ln\Lambda}\Big(\frac{m_0}{\alpha_0}\Big) = -\frac{N_f}{\pi}\frac{d}{d\ln\Lambda}\Big(\frac{m_0}{\alpha_0} \Big) \alpha_0^2 \frac{\partial}{\partial \alpha_0} \ln G(\alpha_0,\Lambda/q) \Big|_{q=0}\nonumber\\
&&\qquad\qquad\qquad\qquad\quad -\frac{N_f}{\pi} m_0\alpha_0 \frac{d\gamma(\alpha_0)}{d\alpha_0}\Big[1 + \frac{N_f}{\pi} \alpha_0^2  \frac{\partial}{\partial\alpha_0}\ln G(\alpha_0,\Lambda/q)\Big|_{q=0}\Big].\qquad
\end{eqnarray}

\noindent
Solving this equation for $d(m_0/\alpha_0)/d\ln\Lambda$, we obtain Eq. (\ref{Exact_Relation}),

\begin{equation}
\frac{d}{d\ln\Lambda}\Big(\frac{m_0}{\alpha_0}\Big) = -\frac{m_0\alpha_0 N_f}{\pi}\cdot \frac{d\gamma(\alpha_0)}{d\alpha_0}.
\end{equation}

\section{$\theta$-dependent renormalization}
\hspace{\parindent}\label{Section_Superfields}

In terms of the $\theta$-dependent coupling constant $A_0 = \alpha_0(1+m_0\theta^2)$ the result obtained for the photino mass renormalization can be combined with the NSVZ $\beta$-function into the superfield NSVZ equation

\begin{equation}
\frac{\beta(A_0)}{A_0^2} = -\frac{d}{d\ln\Lambda}\Big(\frac{1}{A_0}\Big) = \frac{N_f}{\pi}\Big[1-\gamma(A_0)\Big],
\end{equation}

\noindent
which agrees with the results of Ref. \cite{ArkaniHamed:1998kj} and with the general arguments based on the Statement presented in Ref. \cite{Avdeev:1997vx}. Moreover, from Eqs. (\ref{Matter_Two_Point_Function}) and (\ref{Expression_For_g}) we obtain

\begin{equation}
\gamma(A_0) = \frac{d\ln G(A_0,\Lambda/p)}{d\ln\Lambda} = - \frac{d\ln{\cal Z}}{d\ln \Lambda}\Big|_{\bar\theta^2=0},
\end{equation}

\noindent
so that the superfield anomalous dimension is defined in terms of the bare coupling constant by the standard equation.

Note that if the RG functions are defined in terms of the renormalized coupling constant superfield, then the NSVZ relation (\ref{Superfield_NSVZ}) is valid only in a certain class of the subtraction schemes. Really, under the $\theta^2$-dependent finite renormalization of the coupling constant superfield $A$ and the matter superfields, $A' = A'(A,\theta^2)$ and ${\cal Z}'\Big|_{\bar\theta^2=0} = \zeta(A,\theta^2)\, {\cal Z}\Big|_{\bar\theta^2 = 0}$, the RG functions (defined in terms of the renormalized coupling constant) are changed as

\begin{equation}
\beta'(A') = \frac{\partial A'}{\partial A}\cdot \beta(A);\qquad
\gamma'(A') = \frac{\partial \ln \zeta}{\partial A}\cdot \beta(A) + \gamma(A).
\end{equation}

\noindent
This implies that, if the RG functions $\beta(A)$ and $\gamma(A)$ satisfy the NSVZ relation (\ref{Superfield_NSVZ}), then

\begin{equation}
\beta'(A') = \frac{\partial A'}{\partial A}\Big(1-\frac{A^2 N_f}{\pi} \frac{\partial\ln \zeta}{\partial A}\Big)^{-1} \cdot \frac{A^2 N_f}{\pi}\Big(1-\gamma'(A')\Big)\Bigg|_{A=A(A',\theta^2)}.
\end{equation}

\noindent
This equation looks exactly as the corresponding equation for the rigid theory, which has first been derived in \cite{Kutasov:2004xu}. In particular, it implies that after finite renormalizations the NSVZ relation is not in general satisfied, see e.g. \cite{Kataev:2013csa}.

In the end of this section we compare our conventions with the ones used in other papers, in particular, in \cite{ArkaniHamed:1998kj,Luty:1999qc}. For simplicity, here we will assume that there is the only set of the Pauli--Villars superfields with the mass $M$. Let us note that the theory regularized by higher covariant derivatives originally contains two dimensionful parameters, $\Lambda$ and $M$. This is very important, e.g., for investigating anomalies \cite{Novikov:1985mf} and the structure of quantum corrections. In order to obtain the regularized theory with a single dimensionful parameter we use Eq. (\ref{M_Lambda_Relation}), which for a single Pauli--Villars mass may be written as $M=\Lambda$. Then, as we have demonstrated above, the coupling constant $A$ receives quantum corrections in all loops. It is important that the one-loop divergences are regularized by the Pauli--Villars determinants. Consequently, they should be proportional to $\ln M/\mu$. This implies that in the NSVZ scheme the coupling constant $A$ runs according to the equation

\begin{equation}\label{NSVZ_Renormalization}
\frac{1}{A} = \frac{1}{A_0} + \frac{N_f}{\pi} \ln \frac{M}{\mu} + \frac{N_f}{\pi}\ln {\cal Z}\Big|_{\bar\theta^2=0} = \frac{1}{A_0} + \frac{N_f}{\pi}\ln \frac{\Lambda}{\mu} + \frac{N_f}{\pi} \ln {\cal Z}\Big|_{\bar\theta^2=0}.
\end{equation}

However, it is possible \cite{ArkaniHamed:1998kj,Luty:1999qc} to use a different relation between the Pauli--Villars masses and the parameter in higher derivative term, namely, ${\cal M} = \Lambda/{\cal Z}_\phi {\cal Z}_\phi^*$. (We use the notation ${\cal M}$ to distinguish this Pauli--Villars mass from $M = \Lambda$.) Really, after the substitution $\phi \to {\cal Z}_\phi \phi$, $\Phi \to {\cal Z}_\phi \Phi$ the factor ${\cal Z}$ will appear in the kinetic term of the matter superfields, while the mass term will be again equal to $\Lambda$.\footnote{Note that in this case the anomalous contributions coming from the integration measure \cite{ArkaniHamed:1997mj} of the usual matter superfields and of the Pauli--Villars superfields cancel each other.} This follows from Eqs. (\ref{PV_Determinants}) and (\ref{Z_Relation}) taking into account that we consider only terms of the first order in $m_0$ and the limit of small soft scalar masses $m_\phi$.

For the prescription ${\cal M} = \Lambda/{\cal Z}_\phi {\cal Z}_\phi^*$ the holomorphic gauge kinetic coefficient should be considered as a non-physical quantity. This coefficient can be written in terms of the chiral superfield ${\cal A}$ analogous to the expression (\ref{Coupling_Superfield}). It is related to the superfield $S$ in \cite{ArkaniHamed:1998kj} by the equation

\begin{equation}
{\cal A}^{-1}(\alpha,2\theta^2) \equiv  8\pi S(\alpha,\theta^2).
\end{equation}

\noindent
Writing for it an equation similar to (\ref{NSVZ_Renormalization}), we obtain that it runs only at one loop \cite{Shifman:1986zi},\footnote{For ${\cal N}=1$ rigid SQED this has been verified by an explicit calculation up to the three-loop order in \cite{Soloshenko:2003nc}.}

\begin{equation}\label{Holomorphic_Running}
\frac{1}{\cal A} = \frac{1}{A_0} + \frac{N_f}{\pi} \ln \frac{\cal M}{\mu} + \frac{N_f}{\pi}\ln {\cal Z}\Big|_{\bar\theta^2=0} = \frac{1}{A_0} + \frac{N_f}{\pi} \ln \frac{\Lambda}{\mu},
\end{equation}

\noindent
because due to Eq. (\ref{Z_Relation}) ${\cal Z}\Big|_{\bar\theta^2 = 0} = {\cal Z}_\phi {\cal Z}_\phi^*\Big|_{\bar\theta^2 = 0}$. The physical quantity in this case is the real superfield $R$ defined in \cite{ArkaniHamed:1998kj}. In our notation it is written as

\begin{equation}
R \equiv S + S^+ + \frac{N_f}{4\pi^2} \ln {\cal Z}.
\end{equation}

\noindent
In this paper we consider only a part of this superfield which does not contain $\bar\theta^2$, see Eq. (\ref{Theta_Substitution}). This part of the superfield $R$ can written as

\begin{equation}
\frac{1}{A(\alpha,\theta^2)} \equiv 4\pi R(\alpha,\theta^2,\bar\theta^2)\Big|_{\bar\theta^2=0}= \frac{1}{\cal A} + \frac{N_f}{\pi} \ln {\cal Z}\Big|_{\bar\theta^2=0}.
\end{equation}

\noindent
Then from Eq. (\ref{Holomorphic_Running}) we obtain

\begin{equation}
\frac{1}{A} = \frac{1}{A_0} + \frac{N_f}{\pi}\ln \frac{\Lambda}{\mu} + \frac{N_f}{\pi} \ln {\cal Z}\Big|_{\bar\theta^2=0}.
\end{equation}

\noindent
Comparing this equation with Eq. (\ref{NSVZ_Renormalization}) we see that in our conventions the superfield $A$ corresponds to $(4\pi R)^{-1}\Big|_{\bar\theta^2=0}$ in the conventions of \cite{ArkaniHamed:1998kj,Luty:1999qc}.

\section{Explicit two-loop calculation}\label{Section_Two_Loop}
\hspace*{\parindent}

In this section we explicitly calculate the function $d(m_0/\alpha_0)/d\ln\Lambda$ in the two-loop approximation and demonstrate that it is given by integrals of double total derivatives. Moreover, we compare the two-loop result for this function with the one-loop result for the anomalous dimension of the matter superfields. For simplicity,
in this section we will use the gauge

\begin{equation}
\frac{\xi_0}{K} = \frac{1}{R}.
\end{equation}

\noindent
Making the calculation we will omit all terms proportional to $(m_0)^n$ with $n\ge 2$, but keep the terms linear in $m_0$. That is why it is convenient to consider terms linear in $m_0$ as vertexes.\footnote{Note that this method of calculation is different from the one used in the previous sections. Consequently, we will be able to verify independently the general arguments described above.} This implies that the propagator of the gauge superfield is the same as in the rigid theory, namely, it is proportional to

\begin{equation}
\frac{e_0^2}{\partial^2 R(\partial^2/\Lambda^2)}\delta^8_{xy},
\end{equation}

\noindent
and the vertex linear in $m_0$ is given by

\begin{eqnarray}\label{Linear_Vertex}
&& \frac{m_0}{32 e_0^2} \int d^4x\,d^4\theta\,\Big(\theta^2 D^a V R(\partial^2/\Lambda^2) \bar D^2 D_a V + \bar\theta^2 \bar D^{\dot a} V R(\partial^2/\Lambda^2) D^2 \bar D_{\dot a} V\qquad\nonumber\\
&&\qquad\qquad\qquad\qquad\qquad\qquad\qquad  + \theta^2\, V R(\partial^2/\Lambda^2) \bar D^2 D^2 V + \bar\theta^2\, V R(\partial^2/\Lambda^2) D^2 \bar D^2 V\Big).\qquad
\end{eqnarray}

\noindent
We will graphically denote this vertex by a cross on a wavy line.

Let us start with the two-point Green function of the matter superfields. For the rigid theory, regularized by higher derivatives, in the one-loop approximation it was calculated, e.g. in \cite{Kazantsev:2014yna}, and can be written as

\begin{equation}\label{One-Loop_G}
G(\alpha_0,\Lambda/p) = 1 - \int \frac{d^4k}{(2\pi)^4} \frac{2e_0^2}{R_k k^2 (k+p)^2} + O(e_0^4),
\end{equation}

\noindent
where $R_k \equiv R(k^2/\Lambda^2)$. Then the one-loop anomalous dimension (defined in terms of the bare coupling constant) is

\begin{equation}\label{Gamma_Bare}
\gamma(\alpha_0) = - \frac{d\ln Z}{d\ln\Lambda} = \frac{d\ln G}{d\ln\Lambda}\Big|_{p=0} = - e_0^2 \int \frac{d^4k}{(2\pi)^4}\frac{d}{d\ln\Lambda}\frac{2}{R_k k^4} + O(e_0^4)
= -\frac{\alpha_0}{\pi} + O(\alpha_0^2).
\end{equation}

\begin{figure}[h]
\begin{picture}(0,1.5)
\put(4.5,0){\includegraphics[scale=0.34]{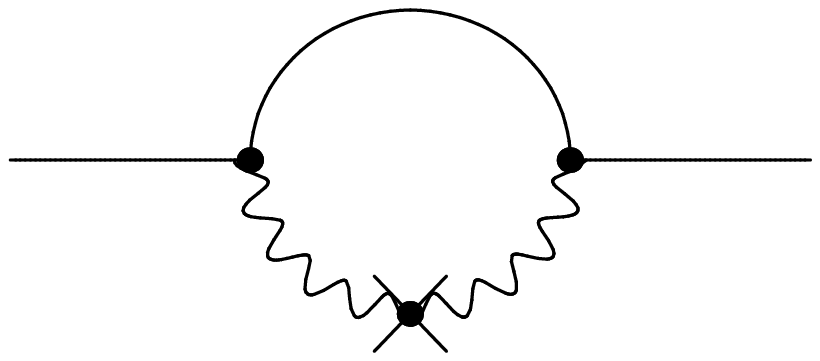}}
\put(8.5,0){\includegraphics[scale=0.33]{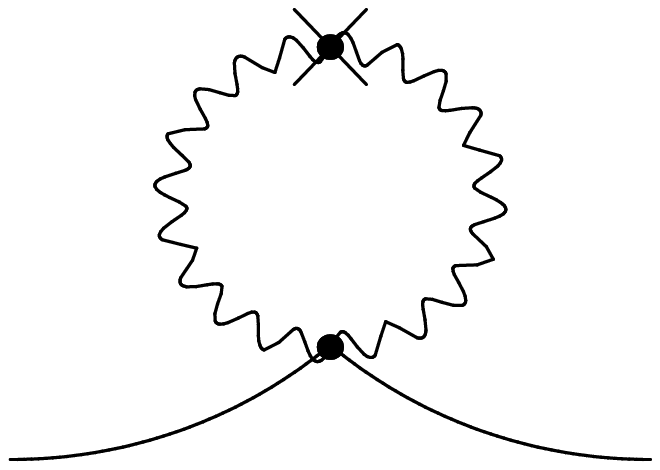}}
\end{picture}
\caption{These diagrams give the $\theta$-dependent part of the two-point Green function of the matter superfields in the lowest order in $m_0$.}\label{Figure_Anomalous_Dimension}
\end{figure}

\noindent
The part of the matter superfield two-point function proportional to $(m_0)^1$ is determined by two diagrams presented in Fig. \ref{Figure_Anomalous_Dimension}. Each of them contains a single vertex (\ref{Linear_Vertex}). We will not consider diagrams with larger numbers of such vertices, because their contributions are proportional to $(m_0)^n$ with $n\ge 2$. After calculating the diagrams presented in Fig. \ref{Figure_Anomalous_Dimension} we obtain the one-loop  contribution to the function $g(\alpha_0,\Lambda/p)$, which is defined by Eq. (\ref{Matter_Two_Point_Function}). The result is written as

\begin{equation}\label{One-Loop_g}
g(\alpha_0,\Lambda/p) = - \int \frac{d^4k}{(2\pi)^4} \frac{2e_0^2}{R_k k^2 (k+p)^2} + O(e_0^4).
\end{equation}

\noindent
Comparing Eqs. (\ref{One-Loop_G}) and (\ref{One-Loop_g}), it is easy to verify correctness of Eq. (\ref{Expression_For_g}), which has essentially been used in deriving the
exact expression for the photino mass renormalization. From Eqs. (\ref{One-Loop_G}) and (\ref{One-Loop_g}) we also conclude that the $\theta$-dependent renormalization
constant for the matter superfield can be chosen in the form

\begin{equation}
{\cal Z}_\phi = 1 + \frac{\alpha}{2\pi}(1+2m_0\theta^2)\ln \frac{\Lambda}{\mu} + \frac{\alpha}{2\pi}(b_1 + 2m_0\theta^2 \widetilde b_1) + O(\alpha^2),
\end{equation}

\noindent
where $b_1$ and $\widetilde b_1$ are finite constants, which fix a subtraction scheme in the considered approximation.

Now, let us verify Eq. (\ref{Result}). The singlet contribution is evidently absent in the considered approximation, so that the first term vanishes. For calculating the
second term, we consider the expression

\begin{equation}
-\frac{i}{8} N_f \frac{d}{d\ln\Lambda} \sum\limits_{I=0}^m c_I \mbox{tr} \Big\langle \bar\theta^2 \psi^2 \ln(\star)\Big\rangle_I,
\end{equation}

\noindent
where $\mbox{tr}$ denotes the usual matrix trace. (Unlike $\mbox{Tr}$, it does not include integration over the superspace.) To calculate it in the considered approximation,
we use Eq. (\ref{Star_Definition}),

\begin{equation}\label{Star_Expansion}
\ln(\star) = - \ln (1-I_0) = I_0 + \frac{1}{2} (I_0)^2 + \ldots = BP + \frac{1}{2} BP BP + \ldots
\end{equation}

\begin{figure}[h]
\begin{picture}(0,1.6)
\put(2.2,0){\includegraphics[scale=0.14]{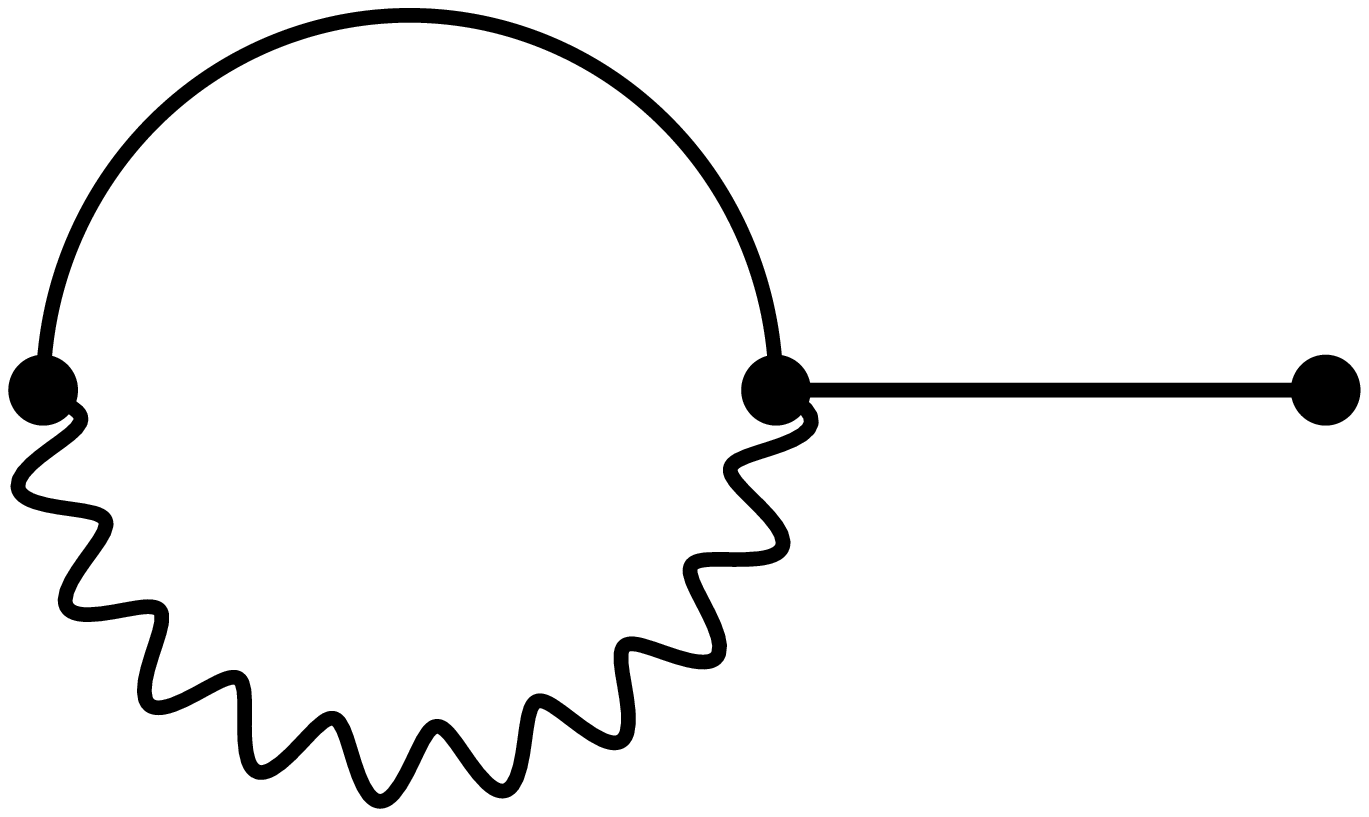}}
\put(5.4,0){\includegraphics[scale=0.14]{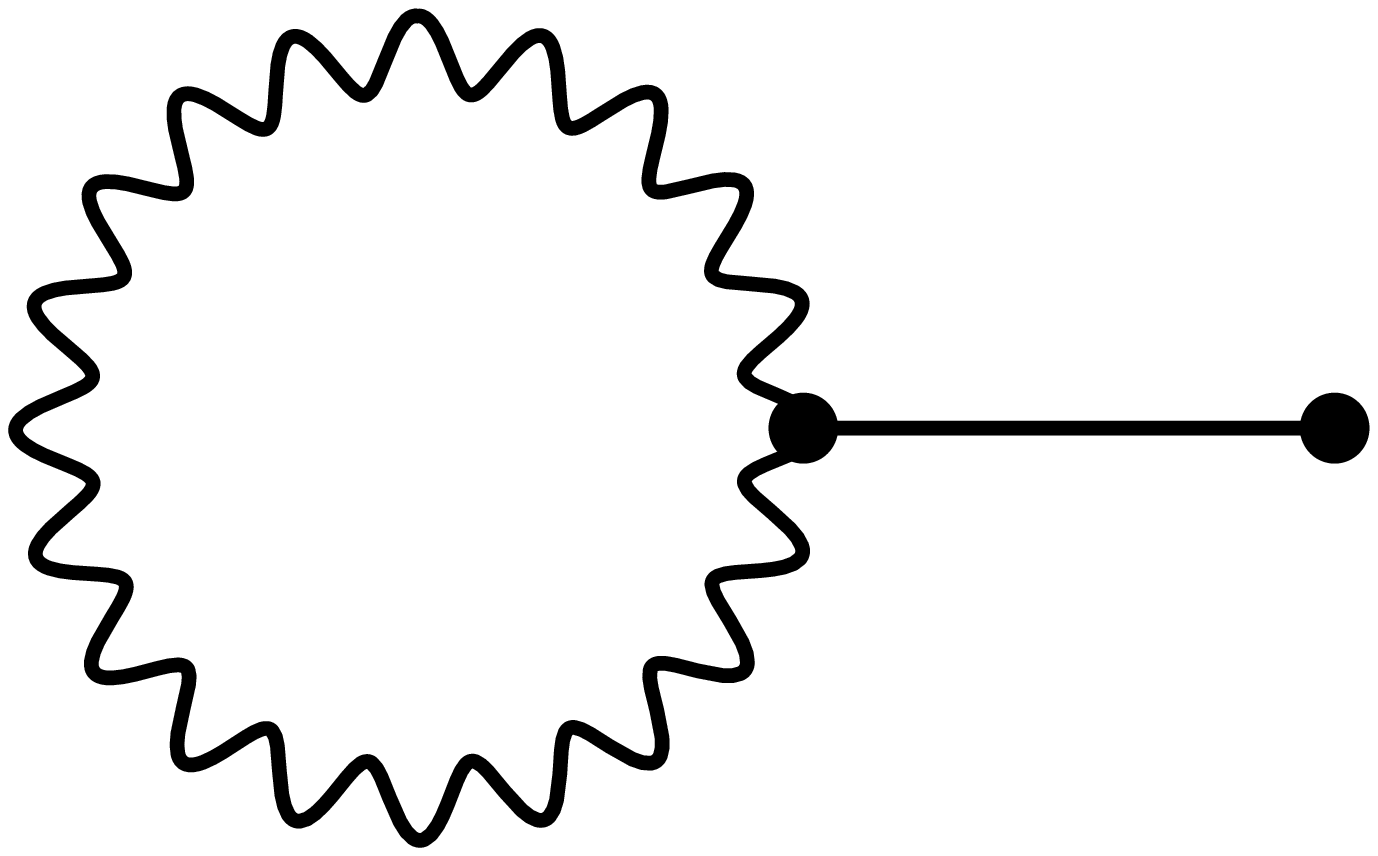}}
\put(8.6,0){\includegraphics[scale=0.14]{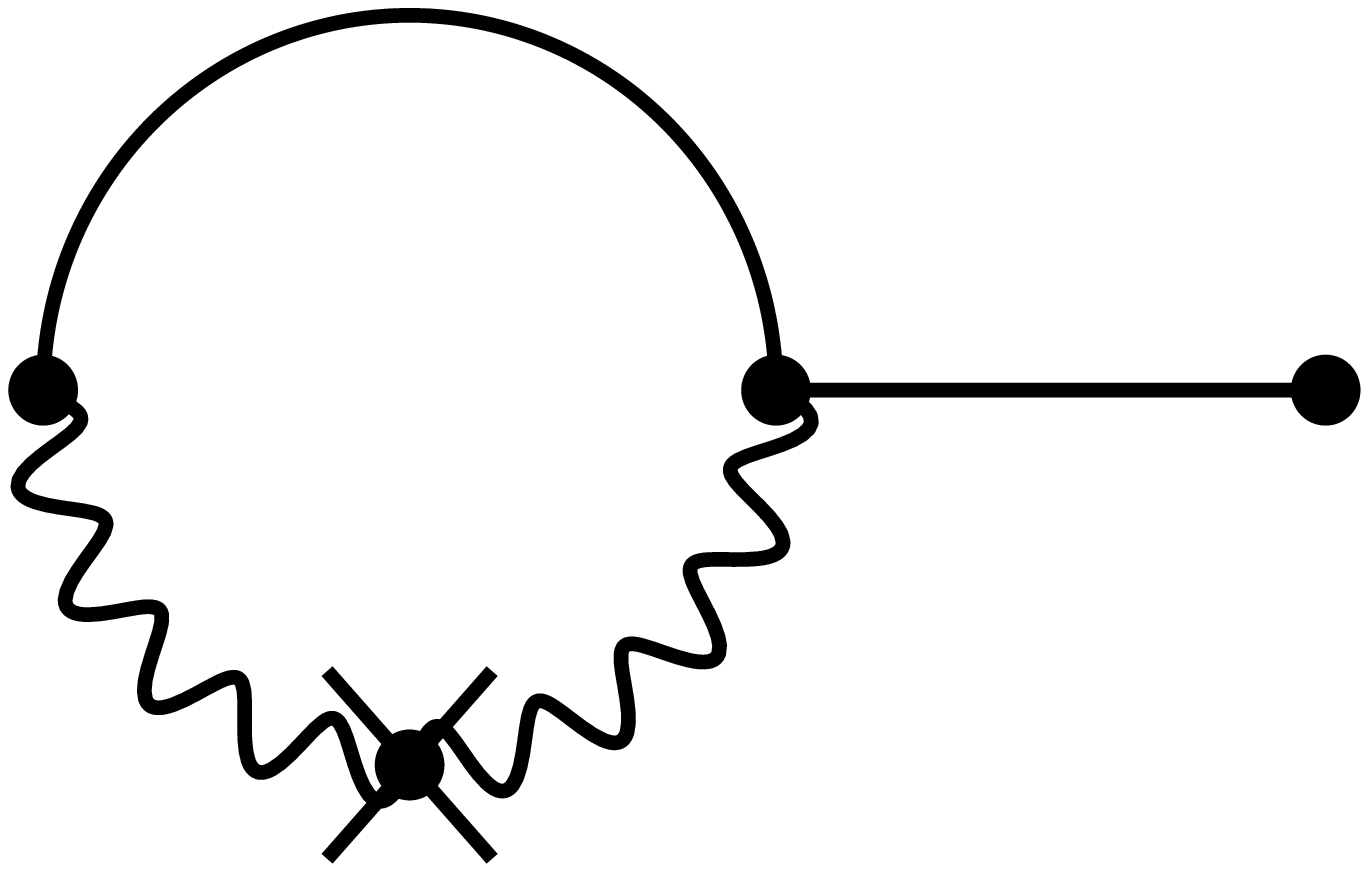}}
\put(11.8,0){\includegraphics[scale=0.14]{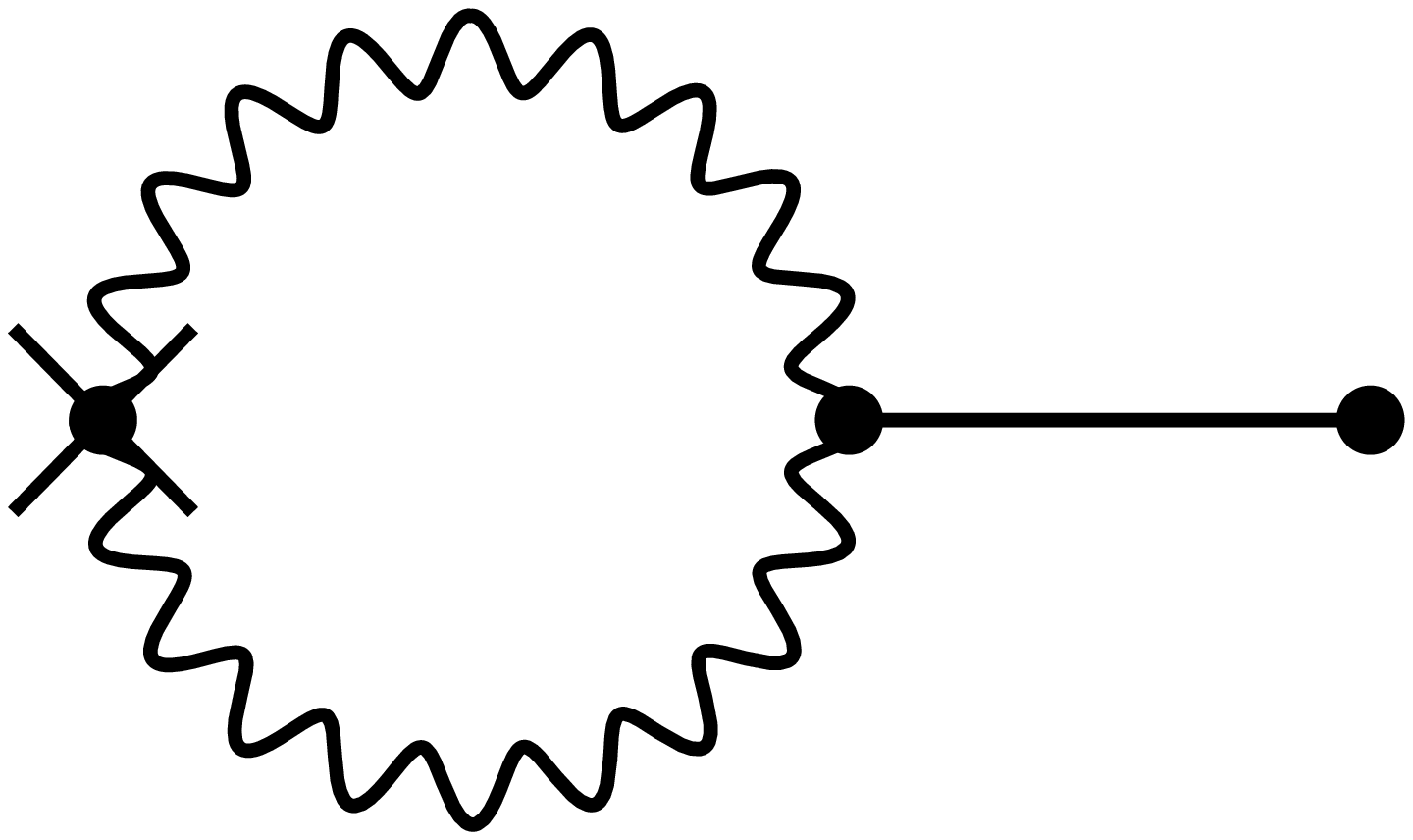}}
\end{picture}
\caption{The supergraphs contributing to the $\mbox{tr}\langle\ln(\star)\rangle$ in the considered approximation.}\label{Figure_Star_Diagrams}
\end{figure}

\noindent
The vertices $B$ contain the gauge superfield $V$. The functional integration encoded in the angular brackets transforms $V$-s into the gauge propagators. The matrices $P$
give propagators of the matter superfields. Then, it is easy to see that from Eq. (\ref{Star_Expansion}) in the lowest approximation we obtain the supergraphs presented in
Fig. \ref{Figure_Star_Diagrams}. (Note that we omit all diagrams proportional to $(m_0)^n$ with $n\ge 2$.) They are similar to the diagrams which contribute to the anomalous
dimension of the matter superfields. Having calculated them we obtained (after the Wick rotation in the Euclidean space)

\begin{eqnarray}
&& \mbox{tr}  \Big\langle \bar\theta^2 \psi^2 \ln(\star)\Big\rangle = -\bar\theta_x^2 \psi_x^2 \int \frac{d^4k}{(2\pi)^4}  \frac{d^4q}{(2\pi)^4} \frac{(1+m_0\theta^2) e_0^2}{k^2 R_k (q^2 + M^2)((q+k)^2 + M^2)} \nonumber\\
&& \times \frac{i}{8} (\bar D^2 D^2 + D^2 \bar D^2) e^{-i q_\alpha (x^\alpha - y^\alpha)} \delta^4(\theta_x-\theta_y).
\end{eqnarray}

\noindent
In the momentum representation $[x^\mu,\ldots]$ is written as $-i\partial/\partial q_\mu$. Therefore, after the Wick rotation in the Euclidean space

\begin{equation}
[x^\mu,[x_\mu, Y]] \to \frac{\partial}{\partial q_\mu} \frac{\partial}{\partial q^\mu} Y.
\end{equation}

\noindent
Using this equation and calculating the remaining integrals in Eq. (\ref{Result}), we can predict the result for the renormalization of the photino mass in the two-loop approximation. Taking into account that the singlet contribution vanishes, we expect that it is described by the function

\begin{eqnarray}\label{Two_Loop_Total_Derivative_Expected}
&& \frac{d}{d\ln\Lambda} \Big(\frac{m_0}{d_m(\alpha_0,\Lambda/p)} - \frac{m_0}{\alpha_0} \Big)\Big|_{p=0; \alpha,m=\mbox{\scriptsize const}} = 16\pi^2 m \alpha N_f \sum\limits_{I=0}^m c_I \int \frac{d^4k}{(2\pi)^4} \frac{d^4q}{(2\pi)^4}\qquad \nonumber\\
&& \times \frac{d}{d\ln\Lambda} \frac{\partial}{\partial q^\mu} \frac{\partial}{\partial q_\mu} \frac{1}{R_k k^2 \left(q^2+M_I^2\right) \left((k+q)^2+M_I^2\right)} + O(\alpha^2).\qquad
\end{eqnarray}

\begin{figure}[h]
\begin{picture}(0,3.7)
\put(0,3.2){$(1)$}
\put(0.1,2.06){\includegraphics[scale=0.34]{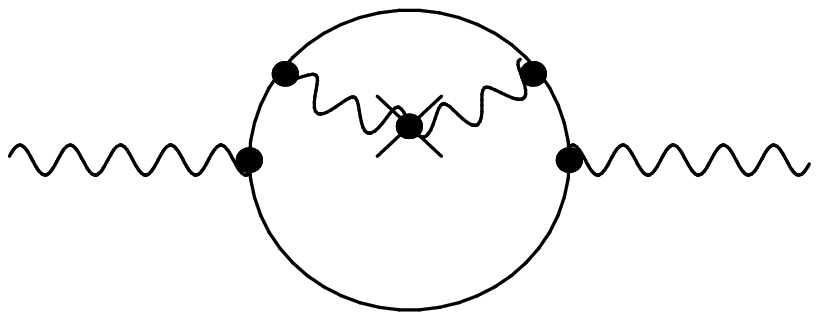}}
\put(3.25,3.2){$(2)$}
\put(3.25,2){\includegraphics[scale=0.34]{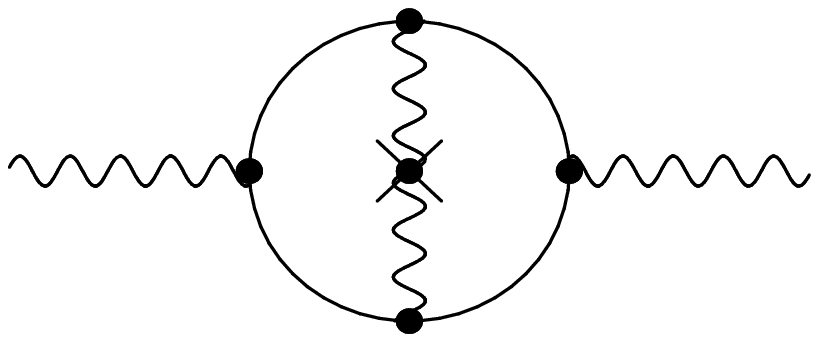}}
\put(6.52,3.2){$(3)$}
\put(6.52,2){\includegraphics[scale=0.34]{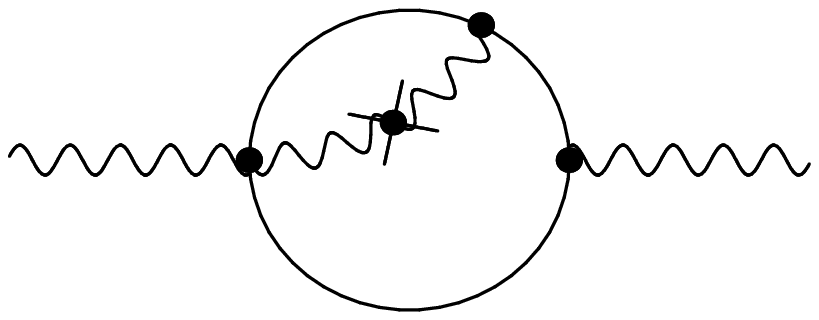}}
\put(9.79,3.2){$(4)$}
\put(9.79,2){\includegraphics[scale=0.34]{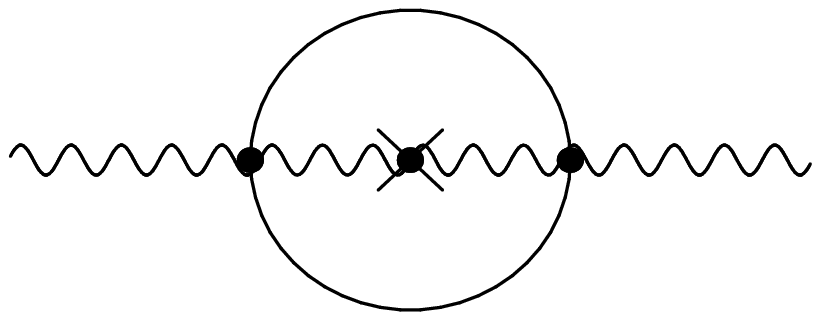}}
\put(13.0,3.2){$(5)$}
\put(13.0,2){\includegraphics[scale=0.34]{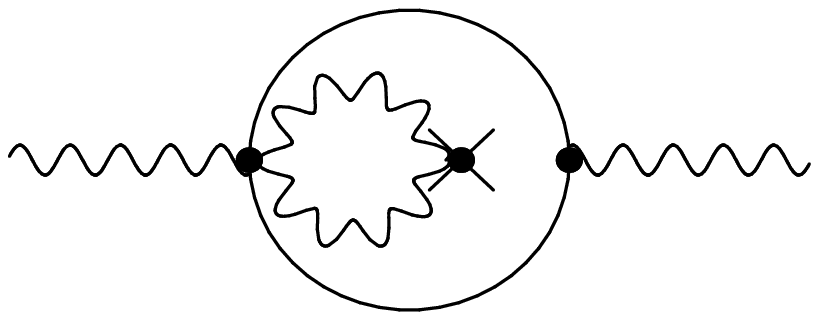}}
\put(0,1.1){$(6)$}
\put(0.1,0){\includegraphics[scale=0.34]{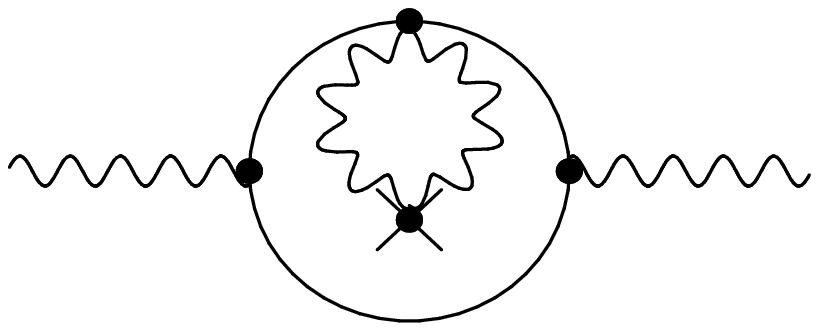}}
\put(3.25,1.1){$(7)$}
\put(3.5,0){\includegraphics[scale=0.34]{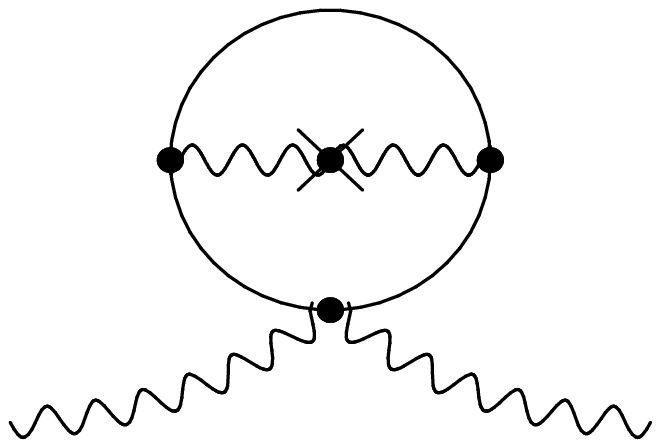}}
\put(6.52,1.1){$(8)$}
\put(6.8,0){\includegraphics[scale=0.34]{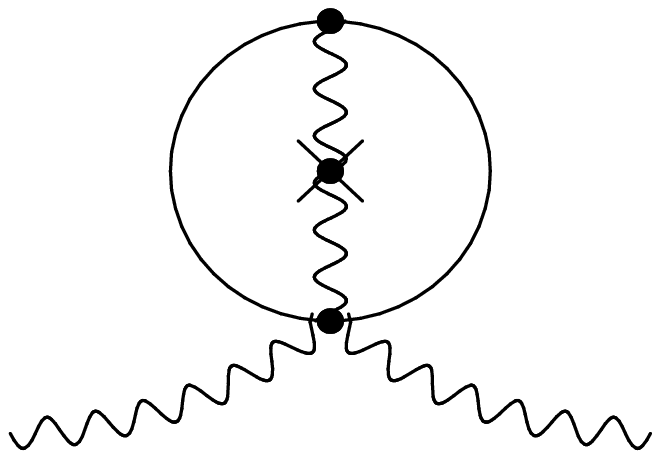}}
\put(9.79,1.1){$(9)$}
\put(10.1,0){\includegraphics[scale=0.33]{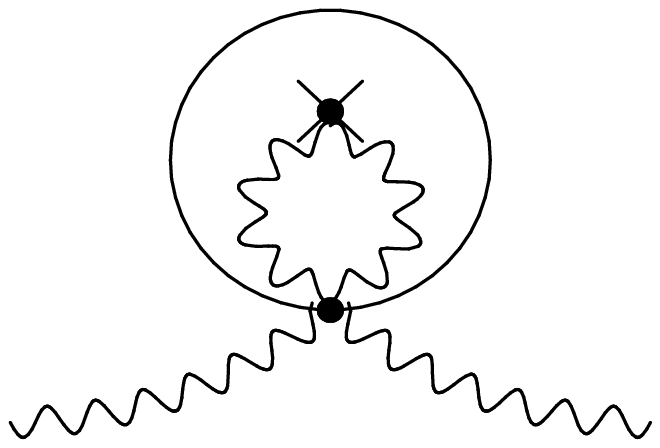}}
\put(13.0,1.1){$(10)$}
\put(13.4,0){\includegraphics[scale=0.33]{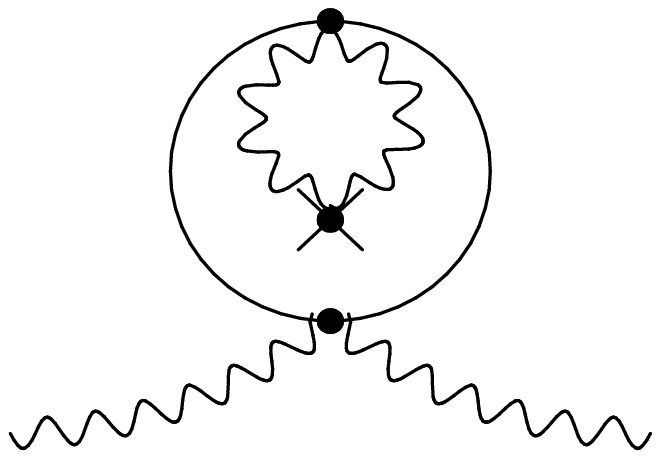}}
\end{picture}
\caption{These diagrams define renormalization of the photino mass in the two-loop approximation in the limit $p\gg m_0$.}\label{Figure_Two_Loop_Diagrams}
\end{figure}

The function $d_m^{-1}$ defined by Eq. (\ref{Gamma2_V}) in the two-loop approximation is given by the sum of the diagrams presented in Fig. \ref{Figure_Two_Loop_Diagrams}. Note that we calculate only terms proportional to first power of $m_0$. The masses inside the function $d_m^{-1}$ (certainly, except for the masses of the Pauli--Villars fields) are set to 0. The expressions which were obtained after calculating the diagrams presented in Fig. \ref{Figure_Two_Loop_Diagrams} are given in Appendix \ref{Appendix_Two_Loop_Diagrams}. These expressions contain 4 different structures, only one surviving in their sum in agreement with the Ward identity. (Cancellation of the terms which do not satisfy the Ward identity can be considered as a non-trivial test of the calculation.)

The sum of the diagrams presented in Fig. \ref{Figure_Two_Loop_Diagrams} is given by Eq. (\ref{Two_Loop_Effective_Action}). Using this equation one can find the function
$d_m^{-1}$ in the considered approximation. After the Wick rotation it can be written as

\begin{eqnarray}
&&\hspace*{-5mm} d_m^{-1}(\alpha_0,\Lambda/p) = \alpha_0^{-1} + 64\pi^2\alpha_0 N_f \sum\limits_{I=0}^m c_I \int \frac{d^4k}{(2\pi)^4} \frac{d^4q}{(2\pi)^4} \frac{1}{R_k k^2}\Bigg(\frac{2 q^\mu (q+k)_\mu -p^2}{\left(q^2+M_I^2\right) \left((q+p)^2+M_I^2\right)}\nonumber\\
&&\hspace*{-5mm} \times \frac{1}{\left((q+k)^2+M_I^2\right) \left((q+k+p)^2+M_I^2\right)} - \frac{4M_I^2}{\left(q^2+M_I^2\right)^2 \left((q+p)^2+M_I^2\right)\left((q+k)^2+M_I^2\right)}\Bigg)\nonumber\\
&&\hspace*{-5mm} + O(\alpha_0^2),\vphantom{\Bigg(}
\end{eqnarray}

\noindent
where $p$ and $k$ are the Euclidean momenta. To calculate the RG function $d(m_0/\alpha_0)/d\ln\Lambda$, we consider the expression (\ref{We_Calculate}). In the considered
approximation it is written as

\begin{eqnarray}\label{Two_Loop_Total_Derivative}
&& \frac{d}{d\ln\Lambda} \Big(\frac{m_0}{d_m(\alpha_0,\Lambda/p)} - \frac{m_0}{\alpha_0} \Big)\Big|_{p=0; \alpha,m=\mbox{\scriptsize const}} = 16\pi^2 m \alpha N_f \sum\limits_{I=0}^m c_I \int \frac{d^4k}{(2\pi)^4} \frac{d^4q}{(2\pi)^4}\qquad \nonumber\\
&& \times \frac{d}{d\ln\Lambda} \frac{\partial}{\partial q^\mu} \frac{\partial}{\partial q_\mu} \frac{1}{R_k k^2 \left(q^2+M_I^2\right) \left((k+q)^2+M_I^2\right)} + O(\alpha^2).\qquad
\end{eqnarray}

\noindent
Therefore, by the explicit calculation we have demonstrated that the integrals defining the function (\ref{Exact_Relation}) are integrals of double total derivatives in the momentum space. Comparing the result with Eq. (\ref{Two_Loop_Total_Derivative_Expected}) we verify that Eq. (\ref{Result}) is valid in the considered approximation.
Thus, the explicit calculation confirms the general argumentation presented in this paper. The integral in Eq. (\ref{Two_Loop_Total_Derivative}) is well-defined due to the derivative with respect to $\ln\Lambda$, which makes this integral convergent in the infrared region. It can be calculated using the equation

\begin{equation}
\int \frac{d^4q}{(2\pi)^4} \frac{q^\mu}{q^4} \frac{\partial}{\partial q^\mu} f(q^2) = - \frac{1}{8\pi^2} f(0),
\end{equation}

\noindent
where $f(q^2)$ is a function with a sufficiently rapid fall-off at infinity. (In the considered case this is ensured by the presence of $R_k$ in the denominator.)
Using this equation it is easy to see that all terms with $I\ne 0$ (which correspond to the diagrams with the Pauli--Villars loops) give vanishing contributions. Therefore, only the term with $I=0$ (for which $c_0=-1$, $M_0=0$) survives:

\begin{eqnarray}
&& \frac{d}{d\ln\Lambda} \Big(\frac{m_0}{\alpha_0} \Big) = -64\pi^2 m \alpha N_f \int \frac{d^4q}{(2\pi)^4} \frac{q^\mu}{q^4} \frac{\partial}{\partial q^\mu}  \frac{d}{d\ln\Lambda} \int \frac{d^4k}{(2\pi)^4} \frac{1}{R_k k^2 (k+q)^2} + O(\alpha^2)\nonumber\\
&& = 8 m_0 \alpha_0 N_f \frac{d}{d\ln\Lambda} \int \frac{d^4k}{(2\pi)^4} \frac{1}{R_k k^4} + O(\alpha_0^2),
\end{eqnarray}

\noindent where we took into account that the differentiation with respect to $\ln\Lambda$ (acting on $R_k$) should be made at fixed values of the renormalized coupling
constant $\alpha$ and the renormalized photino mass $m$. Comparing this expression with the one-loop anomalous dimension of the matter superfields in the rigid theory
calculated with the considered regularization, which is given by Eq. (\ref{Gamma_Bare}), we obtain

\begin{equation}
\frac{d}{d\ln\Lambda} \Big(\frac{m_0}{\alpha_0}\Big) = -\frac{m_0\alpha_0 N_f}{\pi}\cdot \frac{d\gamma(\alpha_0)}{d\alpha_0} + O(\alpha_0^2) = \frac{m_0\alpha_0 N_f}{\pi^2} + O(\alpha_0^2).
\end{equation}

\noindent
Therefore, by the explicit calculation we have verified that Eq. (\ref{Exact_Relation}) is really valid in the two-loop approximation due to factorization of the corresponding integrals into integrals of double total derivatives. Moreover, we have demonstrated that the exact equation (\ref{Result}) is also valid in the considered approximation. Certainly, in the considered approximation the relation (\ref{Soft_Mass_Renormalized}) is scheme-independent (or, similarly, Eq. (\ref{Exact_Relation}) is regularization-independent). That is why it would be more interesting to consider the next order of the perturbation theory. However, the argumentation of this paper is similar to the one presented in \cite{Stepanyantz:2011jy} for the rigid theory, which was checked by the explicit three-loop calculation. Thus, we believe that in the next approximation no principal differences appear.

\section{Conclusion}
\hspace*{\parindent}

In this paper we have derived the exact expression for the anomalous dimension of the photino mass in softly broken ${\cal N}=1$ SQED with $N_f$ flavors by direct summation of supergraphs. The result, which was first proposed in \cite{Hisano:1997ua}, relating the anomalous dimensions of the photino mass and of the matter superfields is obtained for the RG functions defined in terms of the bare coupling constant in the case of using the higher derivative regularization independently of the subtraction scheme. It follows from the fact that all integrals which determine the renormalization of the photino mass are integrals of double total derivatives in the momentum space. This statement is proved in all orders. The integrals of double total derivatives do not vanish due to singularities of the integrands, which can be summed exactly. This sum gives the expression in the right hand side of Eq. (\ref{Exact_Relation}).

The general results obtained in this paper have been verified by the explicit two-loop calculation. This calculation demonstrates that the integrals defining renormalization of the photino mass are really integrals of double total derivatives and coincide with the prediction of the exact Eq. (\ref{Result}).

The RG functions defined in terms of the renormalized coupling constant depend on the subtraction scheme, and, therefore, the relation (\ref{Soft_Mass_Renormalized}) written in terms of the renormalized coupling constant is valid only in a special subtraction scheme. (In this paper we demonstrate the scheme dependence of this equation even in the formalism when the coupling constant is considered as a $\theta$-dependent superfield for incorporating the soft breaking effects.) We believe that, for the theory regularized by higher derivatives, this scheme can be constructed on the base of Eq. (\ref{Exact_Relation}) similarly to the case of the rigid theory considered in \cite{Kataev:2013eta}. Presumably, it can be obtained by imposing boundary conditions analogous to the ones constructed in \cite{Kataev:2013eta} to the $\theta$-dependent coupling constant (\ref{Coupling_Superfield}) and to the matter renormalization constant.

\bigskip

\section*{Acknowledgments}
\hspace*{\parindent}

The K.S. work was supported by the Russian Foundation for Basic Research, grant No. 14-01-00695.

\appendix

\section{Appendix}\label{Appendix_Two_Loop_Diagrams}
\hspace*{\parindent}

Here we present explicit expressions for the diagrams presented in Fig. \ref{Figure_Two_Loop_Diagrams}:

\begin{eqnarray}
&& (5) = (6) = (9) = (10) = 0;\vphantom{\frac{1}{2}}\\
&& (1) = m_0 e_0^2 N_f \sum\limits_{I=0}^m c_I \int \frac{d^4p}{(2\pi)^4} \frac{d^4k}{(2\pi)^4} \frac{d^4q}{(2\pi)^4}\int d^4\theta\, \theta^2 \frac{1}{R_k k^2 ((q+p)^2 - M_I^2) ((q+k)^2 - M_I^2)} \nonumber\\
&& \times \frac{1}{(q^2-M_I^2)}\Bigg\{ \frac{1}{4} \bar D^2 \bm{V}(p,\theta) D^2 \bm{V}(-p,\theta)
- \frac{M_I^2}{2(q^2-M_I^2)} D^a \bm{V}(p,\theta) \bar D^2 D_a \bm{V}(-p,\theta)\nonumber\\
&& + 2(q+p)_\mu (\gamma^\mu)^{a\dot b} \bar D_{\dot b} \bm{V}(p,\theta) D_a \bm{V}(-p,\theta)
+ \bm{V}(p,\theta) \bm{V}(-p,\theta) \frac{4 q^2}{(q^2-M_I^2)} \Big((q+p)^2 -M_I^2 \Big)  \Bigg\}\nonumber\\
&& +\mbox{c.c.};\vphantom{\frac{1}{2}}\\
&& (2) = m_0 e_0^2 N_f \sum\limits_{I=0}^m c_I  \int \frac{d^4p}{(2\pi)^4} \frac{d^4k}{(2\pi)^4} \frac{d^4q}{(2\pi)^4}\int d^4\theta\, \theta^2 \frac{1}{R_k k^4 (q^2-M_I^2)((q+p)^2-M_I^2)}\nonumber\\
&&\times \Bigg\{ \frac{1}{4} \bar D^2 \bm{V}(p,\theta) D^2 \bm{V}(-p,\theta) + \frac{1}{4} D^a \bm{V}(p,\theta) \bar D^2 D_a \bm{V}(-p,\theta) \Bigg(1 -
\frac{2k^2}{((q+k)^2-M_I^2)}
\nonumber\\
&& + \frac{k^2(k^2+p^2-2M_I^2)}{2((q+k)^2-M_I^2)((q+k+p)^2-M_I^2)}\Bigg)  + \bm{V}(p,\theta) \bm{V}(-p,\theta) \frac{2(k-p)^2 (q^2-M_I^2)}{((q+k)^2-M_I^2)} \nonumber\\
&& - 2(\gamma^\mu)^{a\dot b} \bar D_{\dot b} \bm{V}(p,\theta) D_a \bm{V}(-p,\theta) \Bigg(\frac{q_\mu k^\alpha (k-p)_\alpha - p_\mu q^\alpha (q+k)_\alpha + k_\mu q^\alpha (q+p)_\alpha}{((q+k)^2-M_I^2)}\nonumber\\
&&  + \frac{M_I^2(k+p)_\mu}{((q+k+p)^2-M_I^2)} \Bigg) \Bigg\}+\mbox{c.c.};\\
&& (3) = m_0 e_0^2 N_f \sum\limits_{I=0}^m c_I \int \frac{d^4p}{(2\pi)^4} \frac{d^4k}{(2\pi)^4} \frac{d^4q}{(2\pi)^4}\int d^4\theta\, \theta^2
\frac{1}{R_k k^4 ((q+p)^2-M_I^2) ((q+k)^2-M_I^2)}\nonumber\\
&& \frac{1}{(q^2-M_I^2)}\Bigg\{ -\frac{1}{4} \bar D^2 \bm{V}(p,\theta) D^2 \bm{V}(-p,\theta)  \Big( q^2 +(q+k)^2 + k^2 - 2 M_I^2\Big)
- \frac{1}{4} D^a \bm{V}(p,\theta)\nonumber\\
&& \times \bar D^2 D_a \bm{V}(-p,\theta) \Big(q^2 + (q+k)^2 -k^2-2M_I^2\Big) + 2 (\gamma^\mu)^{a \dot b} \bar D_{\dot b} \bm{V}(p,\theta) D_a \bm{V}(-p,\theta) \Big(- q_\mu k_\alpha p^\alpha\nonumber\\
&& - p_\mu (2q^2 + q_\alpha k^\alpha + k^2-2M_I^2) + k_\mu (2q^2 + q^\alpha p_\alpha-2M_I^2) \Big)
- 4 \bm{V}(p,\theta) \bm{V}(-p,\theta) \Big((k-p)^2
\vphantom{\frac{1}{2}}\nonumber\\
&& \times (q^2 - M_I^2) + k^2 ((q+p)^2 -M_I^2)\Big)\Bigg\} + \mbox{c.c.};\\
&& (4) = m_0 e_0^2 N_f \sum\limits_{I=0}^m c_I \int \frac{d^4p}{(2\pi)^4} \frac{d^4k}{(2\pi)^4} \frac{d^4q}{(2\pi)^4}\int d^4\theta\, \theta^2 \frac{1}{R_k k^4 (q^2-M_I^2) ((q+k+p)^2-M_I^2)}\nonumber\\
&& \times \Bigg\{\frac{1}{4} \bar D^2 \bm{V}(p,\theta) D^2 \bm{V}(-p,\theta) + \frac{1}{4} D^a \bm{V}(p,\theta) \bar D^2 D_a \bm{V}(-p,\theta) +2 (k+p)_\mu (\gamma^\mu)^{a \dot b} \bar D_{\dot b} \bm{V}(p,\theta)\nonumber\\
&&\times D_a V(-p,\theta) +2 \bm{V}(p,\theta) \bm{V}(-p,\theta) (k+p)^2 \Bigg\} + \mbox{c.c.};\\
&& (7) = - m_0 e_0^2 N_f \sum\limits_{I=0}^m c_I \int \frac{d^4p}{(2\pi)^4} \frac{d^4k}{(2\pi)^4} \frac{d^4q}{(2\pi)^4}\int d^4\theta\, \theta^2\nonumber\\
&&\qquad\qquad\qquad\qquad\qquad\quad \times \bm{V}(p,\theta) \bm{V}(-p,\theta) \frac{2(q^2+ M_I^2)}{R_k k^2 (q^2 - M_I^2)^2 ((q+k)^2- M_I^2)} + \mbox{c.c.};\\
&& (8) = m_0 e_0^2 N_f \sum\limits_{I=0}^m c_I \int \frac{d^4p}{(2\pi)^4} \frac{d^4k}{(2\pi)^4} \frac{d^4q}{(2\pi)^4}\int d^4\theta\, \theta^2\nonumber\\
&&\qquad\qquad\qquad\qquad\qquad\quad \times \bm{V}(p,\theta) \bm{V}(-p,\theta) \frac{2}{R_k k^2 (q^2 - M_I^2) ((q+k)^2-M_I^2)} + \mbox{c.c.}
\end{eqnarray}

\noindent
We see that these expressions contain 4 different $V$-structures. However, the only combination $D^a \bm{V} \bar D^2 D_a \bm{V}$ is transversal. We have verified that
for the other 3 structures the contributions coming from various diagrams cancel each other. Then, after some transformations, the result (for the contribution of the considered diagrams into the effective action) can be written as

\begin{eqnarray}\label{Two_Loop_Effective_Action}
&& - N_f \sum\limits_{I=0}^m c_I \int \frac{d^4p}{(2\pi)^4} d^4\theta\, \Big(m_0 \theta^2 D^a \bm{V}(p,\theta) \bar D^2 D_a \bm{V}(-p,\theta) +\mbox{c.c.}\Big)\int \frac{d^4k}{(2\pi)^4} \frac{d^4q}{(2\pi)^4} \qquad\nonumber\\
&& \times \frac{e_0^2}{8 R_k k^2} \Bigg(\frac{2 q^\mu (q+k)_\mu -p^2}{\left(q^2-M_I^2\right) \left((q+p)^2-M_I^2\right) \left((q+k)^2-M_I^2\right) \left((q+k+p)^2-M_I^2\right)}\nonumber\\
&& + \frac{4 M_I^2}{\left(q^2-M_I^2\right)^2 \left((q+p)^2-M_I^2\right)\left((q+k)^2-M_I^2\right)}\Bigg).\qquad
\end{eqnarray}


\end{document}